\documentclass[prd, twocolumn, superscriptaddress,floatfix, nofootinbib, preprintnumbers]{revtex4-2}
\pdfoutput=1
\usepackage[utf8]{inputenc}
\usepackage{hyperref}
\hypersetup{colorlinks=true,citecolor=blue}
\usepackage{amssymb}
\usepackage{amsmath}
\usepackage{graphicx}
\usepackage{footmisc}
\usepackage{url}
\usepackage{multirow}
\usepackage{makecell}
\usepackage{enumerate}
\usepackage{hhline}
\usepackage{comment}
\usepackage{nicefrac}
\usepackage{siunitx}
\usepackage{xcolor}
\usepackage{placeins}

\def\beq{\begin{equation}}
\def\eeq{\end{equation}}
\newcommand{\bea}{\begin{eqnarray}\begin{aligned}}
\newcommand{\eea}{\end{aligned}\end{eqnarray}}

\newcommand{\lacathode}{LaC\textsc{athode}}

\begin{document}

\title{Back to the Roots:\\ Tree-Based Algorithms for Weakly Supervised Anomaly Detection}

\author{Thorben Finke}
\email{finke@physik.rwth-aachen.de}
\affiliation{Institut f\"{u}r Theoretische Teilchenphysik und Kosmologie, RWTH Aachen, 52074 Aachen, Germany}

\author{Marie Hein}
\email{marie.hein@rwth-aachen.de}
\affiliation{Institut f\"{u}r Theoretische Teilchenphysik und Kosmologie, RWTH Aachen, 52074 Aachen, Germany}

\author{Gregor Kasieczka}
\email{gregor.kasieczka@uni-hamburg.de}
\affiliation{Institut f\"{u}r Experimentalphysik, Universit\"{a}t Hamburg, 22761 Hamburg, Germany}
\affiliation{Center for Data and Computing in Natural Sciences (CDCS), 22607 Hamburg, Germany}

\author{Michael Kr\"{a}mer}
\email{mkraemer@physik.rwth-aachen.de}
\affiliation{Institut f\"{u}r Theoretische Teilchenphysik und Kosmologie, RWTH Aachen, 52074 Aachen, Germany}

\author{Alexander M\"{u}ck}
\email{mueck@physik.rwth-aachen.de}
\affiliation{Institut f\"{u}r Theoretische Teilchenphysik und Kosmologie, RWTH Aachen, 52074 Aachen, Germany}

\author{Parada Prangchaikul}
\affiliation{Institut f\"{u}r Experimentalphysik, Universit\"{a}t Hamburg, 22761 Hamburg, Germany}

\author{Tobias Quadfasel}
\email{tobias.quadfasel@uni-hamburg.de}
\affiliation{Institut f\"{u}r Experimentalphysik, Universit\"{a}t Hamburg, 22761 Hamburg, Germany}

\author{David Shih}
\email{shih@physics.rutgers.edu}
\affiliation{NHETC, Dept.\ of Physics and Astronomy, Rutgers University, Piscataway, NJ 08854, USA}

\author{Manuel Sommerhalder}
\email{manuel.sommerhalder@uni-hamburg.de}
\affiliation{Institut f\"{u}r Experimentalphysik, Universit\"{a}t Hamburg, 22761 Hamburg, Germany}

\begin{abstract}
Weakly supervised methods have emerged as a powerful tool for model-agnostic anomaly detection at the Large Hadron Collider (LHC). While these methods have shown remarkable performance on specific signatures such as di-jet resonances, their application in a more model-agnostic manner requires dealing with a larger number of potentially noisy input features. In this paper, we show that using boosted decision trees as classifiers in weakly supervised anomaly detection gives superior performance compared to deep neural networks. Boosted decision trees are well known for their effectiveness in tabular data analysis. Our results show that they not only offer significantly faster training and evaluation times, but they are also robust to a large number of noisy input features. By using advanced gradient boosted decision trees in combination with ensembling techniques and an extended set of features, we significantly improve the performance of weakly supervised methods for anomaly detection at the LHC. This advance is a crucial step towards a more model-agnostic search for new physics.
\end{abstract}

\maketitle

\section{Introduction}
The search for new physics at the Large Hadron Collider (LHC) requires powerful and model-agnostic anomaly detection methods~\cite{Kasieczka:2021xcg,Aarrestad:2021oeb}.
Weakly supervised approaches~\cite{Metodiev:2017vrx,Collins:2018epr,Collins:2019jip,Nachman:2020lpy,Andreassen:2020nkr,1815227,Hallin:2021wme,Raine:2022hht,Hallin:2022eoq,Golling:2022nkl,Golling:2023yjq}
have proven effective in identifying specific signatures, such as di-jet resonances, where signal and background can be distinguished using a limited set of hand-crafted features. 
In these approaches,
a classifier is generally trained to distinguish a signal region from a background-dominated control region or background template. In interesting applications, the signal fraction in the signal region is small, and the classifier needs to identify a small number of signal events in an overwhelming background using only the labels for the signal region and background template. Thus, even a relatively simple supervised classification problem becomes increasingly challenging in the weakly supervised setting as the signal fraction decreases. Extending these methods to a wider range of new physics models without prior knowledge of the essential discriminative features requires handling a larger number of potentially noisy input features. This is a significant challenge as the performance of traditional deep neural networks can degrade under such conditions.

In this paper we explore the use of boosted decision trees (BDTs) as classifiers in weakly supervised anomaly detection at the LHC. Boosted decision trees are known for their strength in analyzing tabular data~\cite{grinsztajn2022treebased}. In particular, on small and medium training sets, BDTs outperform deep learning methods, see for example the reviews~\cite{9998482, grinsztajn2022treebased} and references therein. In the case of weakly supervised anomaly detection, only the signal events separate the two classes. Therefore, while the total training set may be large, the small number of signal data may favor BDTs. In addition, BDTs are generally less affected by uninformative features \cite{grinsztajn2022treebased} (see also Appendix \ref{app:uninformative}). In a high-dimensional input feature space for truly model-agnostic anomaly detection, many of the input features will inevitably be uninformative for a given signal model. As such, robustness to uninformative features is an important property of a model-agnostic method. This robustness and the efficient training and evaluation times make BDTs an interesting alternative to deep neural networks. 

An important aspect of improving performance in weakly supervised learning is ensembling, which harnesses the collective power of multiple classifiers to achieve improved classification accuracy. Although ensembling can be applied to various machine learning methods, the fast training and evaluation time of BDTs allows ensembling to be used most efficiently. 

We study the potential superiority of boosted decision trees over deep neural networks for weakly supervised anomaly detection using modern gradient boosting techniques and ensembling methods. Having obtained classifiers that scale well to a larger number of input features, we investigate whether weak supervision techniques can take advantage of additional quantities describing the substructure of jets. 

The remainder of this paper is structured as follows: 
Section~\ref{sec:method} describes weak anomaly detection and modern boosted decision trees;
Section~\ref{sec:data} introduces the LHC Olympics (LHCO) benchmark dataset used;
Section~\ref{sec:results} discusses the empirical results of our studies;
and Section~\ref{sec:conclusions} presents a summary and outlook for future work. In Appendix~\ref{app:bdt_hyper} we compare the performance of different BDT architectures, in Appendix~\ref{app:uninformative} we analyze the robustness of BDTs with respect to noisy features, and in Appendix~\ref{app:ensemble} we discuss the impact of ensembling in more detail. 

\section{Methods}
\label{sec:method}

\subsection{Optimal anomaly score and its approximations}
\label{sec:iad}

According to the Neyman-Pearson lemma~\cite{Neyman:1933wgr}, 
the most powerful model-agnostic anomaly score is the likelihood ratio between data and background events
\begin{equation}\label{eq:optimalscore}
R_{\rm optimal}(\boldsymbol{x})=\frac{p_{\mathrm{data}}(\boldsymbol{x})}{p_{\mathrm{bg}}(\boldsymbol{x})}\,,
\end{equation}
where $\boldsymbol{x}$ is a set of features that describe the events (e.g.\ kinematics, high-level variables etc.). This is monotonic with the signal-to-background likelihood ratio for any, not necessarily specified, signal model. We will refer to (\ref{eq:optimalscore}) as the {\it optimal anomaly score}. 

In practice, in any realistic high energy physics context, it is not possible to achieve the optimal anomaly score, for several reasons. First, the true background density is generally unknown---our simulations of the Standard Model and the detector response are imperfect, so at best we have an approximation to $p_{\rm bg}({\boldsymbol x})$. Furthermore, the  densities themselves are intractable and at best we have samples from them. Finally, the data density is also unknown and must also be approximated from samples. 

Existing methods~\cite{Metodiev:2017vrx,Collins:2018epr,Collins:2019jip,Nachman:2020lpy,Andreassen:2020nkr,1815227,Hallin:2021wme,Raine:2022hht,Hallin:2022eoq,Golling:2022nkl,Golling:2023yjq} attempt to approximate the optimal anomaly score through binary classifiers that are trained to distinguish between samples drawn from the events within a signal region (SR) and samples drawn from a data-driven background template. To date, this classification task has  been exclusively performed using deep neural networks. This work will instead focus on the benefits of using boosted decision trees.

So far, methods have primarily focused on the case of {\it resonant anomaly detection}, where the SR is a window in a resonant variable such as dijet invariant mass, and the background template is derived somehow from neighboring control regions. However, the idea of the optimal anomaly score and approximating it with a classifier is more general and not necessarily limited to resonant anomaly detection, see e.g.\ \cite{Finke:2022lsu}.

In order to separate out the different approaches to deriving background templates from the performance of the binary classifier itself---the issue of interest in this work---we will focus on a version of the optimal anomaly score, previously termed the {\it idealized anomaly detector} (IAD) in \cite{Hallin:2021wme}, which is a binary classifier trained on SR data and a {\it perfectly modeled} background template. In other words, we presume that the background events in the SR and the background template events are drawn from the exact same distribution. Focusing on the IAD in our work ensures that any relative improvement in significance between approaches is solely due to an improvement in the classification task itself. 

\subsection{Boosted decision trees}

\label{sec:bdts}

Gradient Boosted Decision Trees (GBDTs) are established as the best performing methods on tabular data. In recent years, LightGBM \cite{Ke:2017lgbm} has become the state of the art, mainly due to its fast training and evaluation times achieved by histogramming the inputs \cite{Carlens:2023state}. We use the \texttt{HistGradientBoostingClassifier} implementation in \texttt{scikit-learn}~\cite{Pedregosa:2011sk}, which is based on LightGBM, because it is easy to use and performs well. We use default hyperparameters unless stated otherwise, i.e.\ a learning rate of $0.1$, a maximum number of leaf nodes per tree of 31, a maximum of 255 bins per feature, and early stopping with a patience of 10 iterations. The maximum number of iterations is increased with respect to the default to 200 to ensure that all training is stopped at the minimum validation loss and not at the maximum number of trees. The GBDT implementation uses subsampling for individual trees such that the predictions can vary for difficult classification tasks like weakly supervised anomaly detection with small signal fractions. 

To further stabilize and improve performance, our classifier uses an ensemble of $N$ independent runs of the BDT by averaging the $N$ predictions. For this ensemble, the split between training and validation is randomized. This results in a further increase in performance compared to a fixed split. Unless otherwise specified, our classifier uses an ensemble of $N=50$ independent runs. 

We compare the performance of our default GBDT implementation with other BDT architectures in Appendix~\ref{app:bdt_hyper}.

\subsection{Neural Network}
\label{sec:nn}
We use the neural network architecture of Ref.~\cite{Hallin:2021wme}, however we have implemented it in Tensorflow \cite{tensorflow2015-whitepaper} using Keras \cite{Chollet:2015keras}. The network is a fully connected neural network with a binary cross-entropy loss. It consists of 3 hidden layers, each containing 64 nodes and using ReLU activation. The output layer employs the softmax activation. Training is conducted over 100 epochs using the Adam \cite{Kingma:2014ad} optimizer, a learning rate of $10^{-3}$, and a batch size of 128. Early stopping with a patience of 10 epochs is used in analogy to the BDT. The NN classifier also applies an ensemble of $N=50$ independent runs with a randomized validation split.

\section{data}
\label{sec:data}
\subsection{The dataset}
\label{sec:dataset}

Our studies are performed using the R\&D dataset~\cite{LHCOdataset} from the LHC Olympics 2020 (LHCO) \cite{Kasieczka:2021xcg}. For the most part, we focus on the original R\&D signal model, $Z'\rightarrow X(\rightarrow qq) Y(\rightarrow qq)$, with masses $m_{Z'}=\SI{3.5}{\tera\electronvolt}$, $m_{X}=\SI{500}{\giga\electronvolt}$ and $m_{Y} = \SI{100}{\giga\electronvolt}$. In this signal model, the boosted dijets have 2-prong substructure. In Section~\ref{subsec:3-prong}, we further investigate an alternative signal model also included with the LHCO R\&D dataset, where the $X$ and $Y$ particles decay to 3-prong substructure, i.e.\ $Z'\rightarrow X(\rightarrow qqq) Y(\rightarrow qqq)$~\cite{LHCOdataset}. 

The dataset contains $\num{1000000}$ QCD dijet events for the background and $\num{100000}$ signal events, both simulated with \texttt{Pythia 8}~\cite{Sjostrand:2006za,Sjostrand:2007gs} and \texttt{Delphes 3.4.1}~\cite{deFavereau:2013fsa}. Reconstructed particles are clustered into jets with \texttt{FastJet}~\cite{Cacciari:2011ma} using the anti-$k_{T}$ algorithm~\cite{Cacciari:2005hq} with a distance parameter of $R=1$. Additionally, a trigger of $p_{\mathrm{T}}>\SI{1.2}{\tera\electronvolt}$ must be passed for all events.

Of the background events, approximately \num{120000} fall in the SR between $\SI{3.3}{\tera\electronvolt}$ and $\SI{3.7}{\tera\electronvolt}$. We also use \num{612858} additional QCD background events in the SR, which can be found here \cite{extraLHCOdataset}, and which have been employed previously in Ref.~\cite{Hallin:2021wme}. These events are used for the 
background template for the IAD, for the supervised classifier, and for the construction of the test dataset. 

Unless otherwise specified, $\num{1000}$ signal events are injected into the $\num{1000000}$ background events of the R\&D dataset. As in Ref.~\cite{Hallin:2021wme}, the same signal events are used in different classifiers unless $S/B$ is varied. This allows for better comparability between BDT and NN results, as the high computational cost of the NN makes it prohibitive to train multiple classifiers for each setup (see also Section~\ref{subsec:metric}). This results in 772 signal events in the SR and corresponds to $S/B = 6 \times 10^{-3}$ and $S/\sqrt{B}=2.2$.

For training and validation, a total of approximately 272k SR background events in the background template are trained against 120k SR data events for the IAD and 54k SR signal events for the supervised classifier. In both cases a 50\,\% validation split is used, which is randomized between different runs as mentioned above. Testing is performed on 340k SR background and 20k SR signal events. 

\subsection{Features}

The features used for training are similar to those used in Ref.~\cite{Hallin:2021wme}. We select the two jets with the highest $p_T$. As features we use the invariant mass of the lighter of the two jets, $m_{J_{1}}$, the difference in jet mass between the two jets, $\Delta m_{J} = m_{J_{2}}-m_{J_{1}}$, and several features based on their n-subjettiness~\cite{Thaler:2010tr,Thaler:2011gf}. The original R\&D dataset consisted of $\tau_{1}$, $\tau_{2}$ and $\tau_{3}$ subjettiness features computed with angular weighting parameter $\beta=1$. In this work, we also computed additional subjettiness features ($\tau_n$ up to $n=9$) for both jets, and using three different values for $\beta$ ($\beta = 0.5$, $\beta = 1$ and $\beta = 2$). Thus, a total of $\num{54}$ subjettiness features are available considering both jets. To investigate the impact of the subjettiness features on the classification performance, we consider various feature sets as listed in Tab.~\ref{tab:features}. The subjettiness ratios are defined as $\tau_{ij} \equiv \tau_i/\tau_j$. 

\begin{table}[t]
\def\arraystretch{1.5}
\begin{tabular}{c|c|c}
Name       &  \# features & Features                                              \\\hline
Baseline   & 4                  & $\{m_{J_1},\,\Delta  m_J,\,\tau_{21}^{\beta=1,J_1},\,\tau_{21}^{\beta=1,J_2}\}$ \\\hline
\multirow{2}{*}{Extended 1} & \multirow{2}{*}{10}                & $\{m_{J_1},\,\Delta  m_J,\,\tau_{N, N-1}^{\beta=1,J_1},\, \tau_{N, N-1}^{\beta=1,J_2}\}$ \\ & & for $2\le N\le 5$                                 \\\hline
\multirow{2}{*}{Extended 2} & \multirow{2}{*}{12}               & $\{m_{J_1},\,\Delta  m_J,\,\tau_{N}^{\beta=1,J_1},\, \tau_{N}^{\beta=1,J_2}\}$ \\ & & for $N\le 5$   \\\hline
\multirow{2}{*}{Extended 3} & \multirow{2}{*}{56}                & $\{m_{J_1},\,\Delta  m_J,\,\tau_{N}^{\beta, J_1},\, \tau_{N}^{\beta, J_2}\}$  \\ & & for $N\le 9$ and $\beta\in\{0.5, 1, 2\}$  \\\hline
\end{tabular}
\caption{Subjettiness feature sets considered for training. Full training feature sets always include $m_{J_1}$ and $\Delta m_J$ as well. Details of the observables are given in the text.\label{tab:features}} 
\end{table}

\subsection{Metric}\label{subsec:metric}

Throughout this work we use the significance improvement characteristic (SIC) as our main metric. We define 
\begin{equation}\label{eq:sic}
\textrm{SIC} = \frac{\epsilon_S}{\sqrt{\epsilon_B}}\,   
\end{equation}
as the ratio of the fraction of correctly identified  signal events $\epsilon_S$ to the square root of the fraction of background events misidentified as signal $\sqrt{\epsilon_B}$. This is an approximation of the expected sensitivity gain over an inclusive search, assuming a dominant Poisson uncertainty in the background event count. For very low background and signal efficiencies, the statistical error of the SIC value increases, causing large fluctuations. As the uncertainty of the background efficiency dominates, a cut-off on the relative statistical error of the background efficiency at 20\% is introduced.

For the BDT, all results show the median and 68\,\% confidence intervals of 10 independent classifiers each employing our ensembling procedure. As training the NNs is computationally intensive, the results of only one NN classifier are shown. However, for the large ensemble of $N=50$ independent trainings per classifier, the variance is expected to be small; this has been verified by examining 10 different NN classifiers with $N=10$ for all results presented here. 

\section{Results} 
\label{sec:results}

\subsection{Stability under noisy features}\label{subsec:noise}

\begin{figure}[ht]
        \centering
        \includegraphics[width=0.5\textwidth,trim=0 0 0 20, clip=]{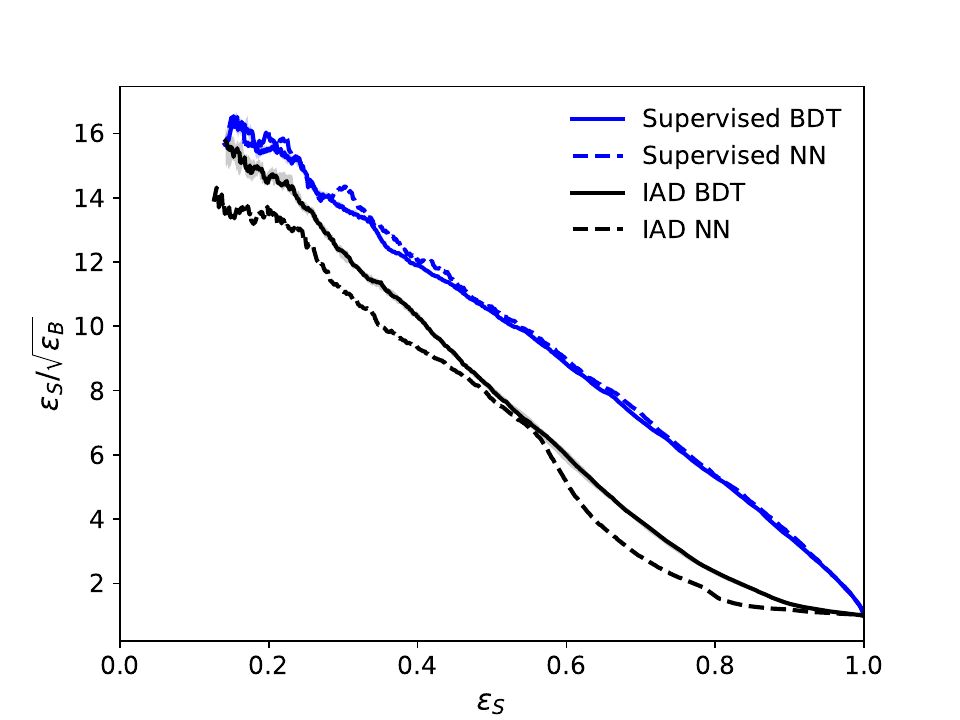}
        \caption{Significance improvement characteristic (SIC) curve, eq.~(\ref{eq:sic}), for the baseline scenario with four features introduced in \cite{Hallin:2021wme}. We compare the idealized anomaly detector (IAD) and a fully supervised setting for the BDT and the NN classifier. } 
        \label{fig:sic-NN-BDT-original-4-features}
\end{figure}

We first investigate our so-called baseline setup which considers the four features $m_{J_{1}}$, $\Delta m_{J}$, $\tau_{21}^{1}$ and $\tau_{21}^{2}$ and is close to the one originally considered in \cite{Hallin:2021wme}. We confirm that BDTs perform at least on par with NNs in this setup as shown in Fig.~\ref{fig:sic-NN-BDT-original-4-features}. The IAD BDT outperforms the IAD NN for all values of signal efficiency and achieves a maximum SIC value of 15. As expected, supervised training using signal and background labels outperforms the weakly supervised IAD, with a difference in maximum achievable SIC of about 2. For the relatively simple supervised classification task, BDT and NN achieve the same performance.

\begin{figure*}[ht]
        \centering
        \includegraphics[width=0.495\textwidth,trim=0 0 0 30, clip=]{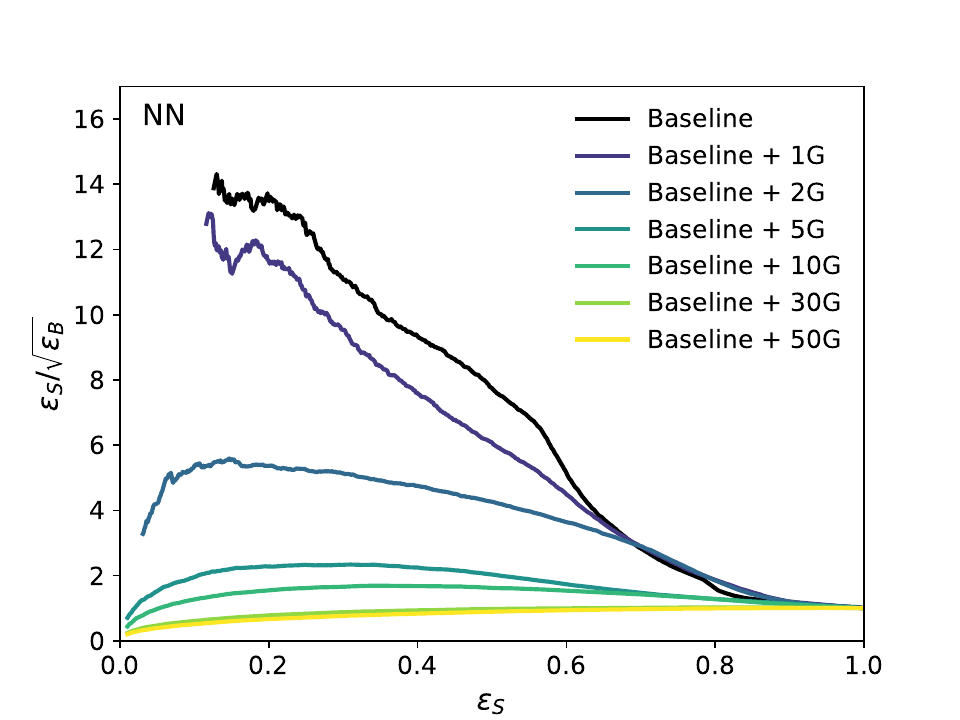}
        \includegraphics[width=0.495\textwidth,trim=0 0 0 30, clip=]{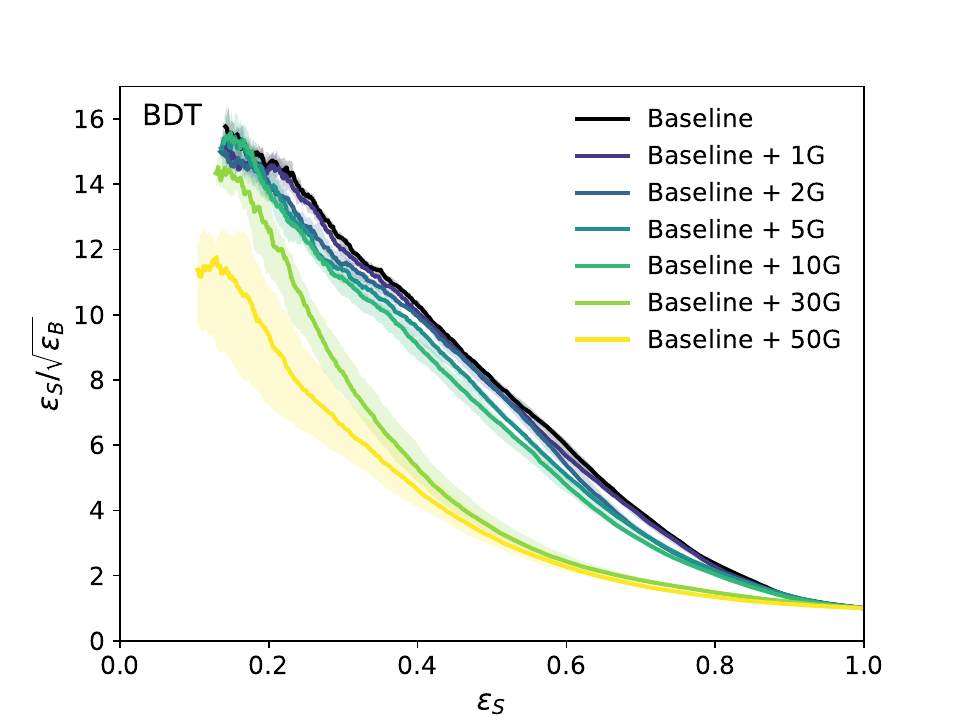}
        \caption{The impact of uninformative features on the NN and BDT classifiers. We show the SIC curves of the IAD NN/BDT classifiers employing the four baseline features and additional 1, 2, 5, 10, 30 and 50 input features drawn from Gaussian noise (left/right panel). For 30 and 50 Gaussian noise features, BDT classifiers with $N=100$ are used instead of the usual $N=50$ ensembling.}
        \label{fig:sic-Gaussian_features}
\end{figure*}

Next, we test the stability of NN- and BDT-based idealized anomaly detection in the presence of noisy features. Since in anomaly detection the potential signal is not known, it is expected that some input features do not carry relevant information for its discrimination. We mimic this by adding additional input features that consist of Gaussian noise with mean zero and a standard deviation of one. The rest of the architecture is unchanged and the noise features are drawn from the {\it same} normal distribution for signal and background. In Fig.~\ref{fig:sic-Gaussian_features} (left panel) we observe a drastic drop in SIC to a maximum value below 4 when only five Gaussian noise (5G) features are included in the idealized neural network classifier. The performance of the BDT IAD, on the other hand, is much more stable against uninformative features. This is illustrated in Fig.~\ref{fig:sic-Gaussian_features} (right panel), where we show that the BDT IAD still performs significantly better than random for up to 50 Gaussian noise features. The robustness of the BDT with respect to uninformative features is analyzed in more detail in Appendix \ref{app:uninformative}. 

Ensembling, as introduced in Section~\ref{sec:bdts}, is  important for the performance and stability of the BDT classifier in the presence of noisy features. For the NN, on the other hand, the improvement due to ensembling is much less significant (and much more expensive to obtain). As the size of the error band suggests, the BDT classifier would even benefit from a larger ensemble with $N>100$ when many noisy features are added. The effect of ensembling is discussed in more detail in Appendix~\ref{app:ensemble}. 

\subsection{Expanding the pool of features}\label{subsec:SIC_features}

\begin{figure*}[ht]
        \centering
        \includegraphics[width=0.495\textwidth,trim=0 0 0 30, clip=]{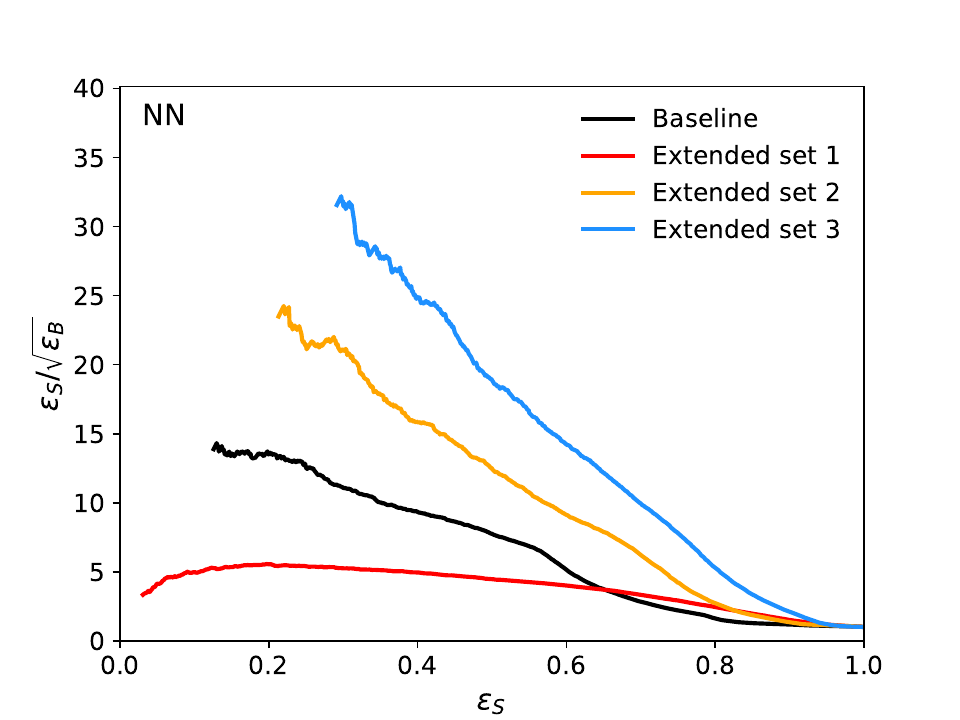}
        \includegraphics[width=0.495\textwidth,trim=0 0 0 30, clip=]{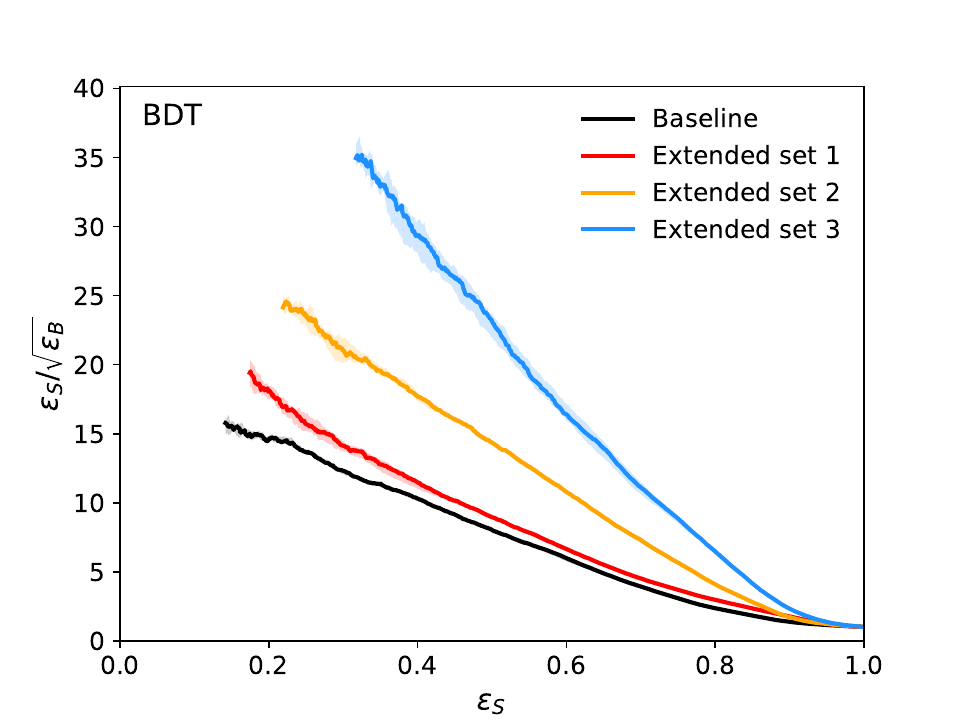}
        \caption{The impact of physics features on the NN and BDT classifiers. We show the SIC curves of the IAD NN with the four baseline features and the extended feature sets based on subjettiness variables as introduced in Tab.~\ref{tab:features} (left panel), and the corresponding results for the IAD BDT (right panel).}   
        \label{fig:sic-extended_sets}
\end{figure*}

Having identified the BDTs as a method that is stable under the inclusion of potentially noisy features, we can explore how additional physics features can improve anomaly detection performance. In Fig.~\ref{fig:sic-extended_sets}, we compare the SIC curves for the IAD in the baseline setup as in \cite{Hallin:2021wme} with the IAD employing various extended feature sets using additional jet substructure information. The extended sets are given in Table~\ref{tab:features}, and include additional subjettiness ratios (extended set 1), individual subjettiness features up to $\tau_5$ (extended set 2), and a set of 54 subjettiness features computed with different angular weighting parameters $\beta$ (extended set 3). We show results for the NN IAD classifier (Fig.~\ref{fig:sic-extended_sets}, left panel) and the BDT IAD classifier (Fig.~\ref{fig:sic-extended_sets}, right panel). 

As expected from the study of Gaussian noise, Section~\ref{subsec:noise}, the NN IAD classifier is very sensitive to the selection of features. The inclusion of additional subjettiness ratios leads to a dramatic reduction in performance, while the inclusion of individual subjettiness features improves the classification properties. 

The BDT IAD classifier, on the other hand, is characterized not only by its higher performance, but also by its stability under the inclusion of distinct features. Additional subjettiness ratios and, in particular, the inclusion of individual subjettiness features lead to a dramatic improvement in performance. Thus, unlike the NN, adding more and more features does not affect the performance of the BDT. 

From our studies in Section~\ref{subsec:noise}, it is clear that adding more and more noisy features will eventually degrade performance. We observed a slight drop in performance for the BDT when we added even more features (beyond extended set 3) based on the  subjettinesses. However, again the NN classifier suffers much more in this case.

\subsection{Dependence on amounts of signal and background }\label{subsec:maxSIC}

So far we have only considered a fixed number of signal and background events in the SR. To detect a potential signal, it is of course not essential whether our classifier reaches ${\rm SIC}=20$ or ${\rm SIC}=40$. Much more important is the minimum signal fraction that is still detected by the classifier for a given signal region size. Therefore, it is important to investigate the potential sensitivity of the classifiers for a range of signal fractions and signal region sizes.

\begin{figure}[ht]
    \centering
    \includegraphics[width=0.5\textwidth,trim=0 0 0 30, clip=]{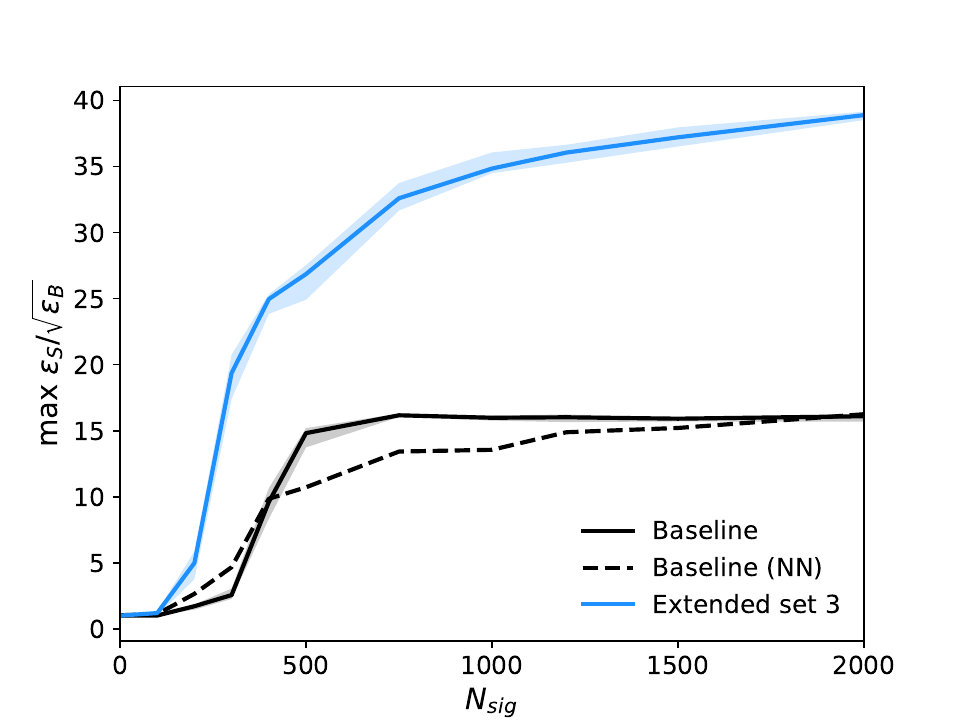}
    \caption{The impact of physics features on the sensitivity improvement. We show the maximum SIC values as a function of the number of signal events in the SR, with the number of background events held fixed. The BDT IAD performance is shown for the baseline feature set and the extended feature set 3. The NN performance is shown for the baseline set for comparison.}
    \label{fig:1d-scan}
\end{figure}

Figure~\ref{fig:1d-scan} shows the maximum SIC value as a function of the number of signal events in the SR for the default fixed number of 120k background events. 
Achieving significance improvement for rarer signals will translate into analysis sensitivity for lower cross-sections. For the baseline set of features, the BDT plateaus at a SIC of 15, which is reached with a minimum of approximately 500 signal events, corresponding to $S/B = 4 \times 10^{-3}$. The same SIC value can be reached with the extended feature set 3 with less than half the number of events, reaching a plateau SIC value of about 40.

\begin{figure}[b]
        \centering
        \includegraphics[width=0.495\textwidth,trim=0 0 20 30, clip=]{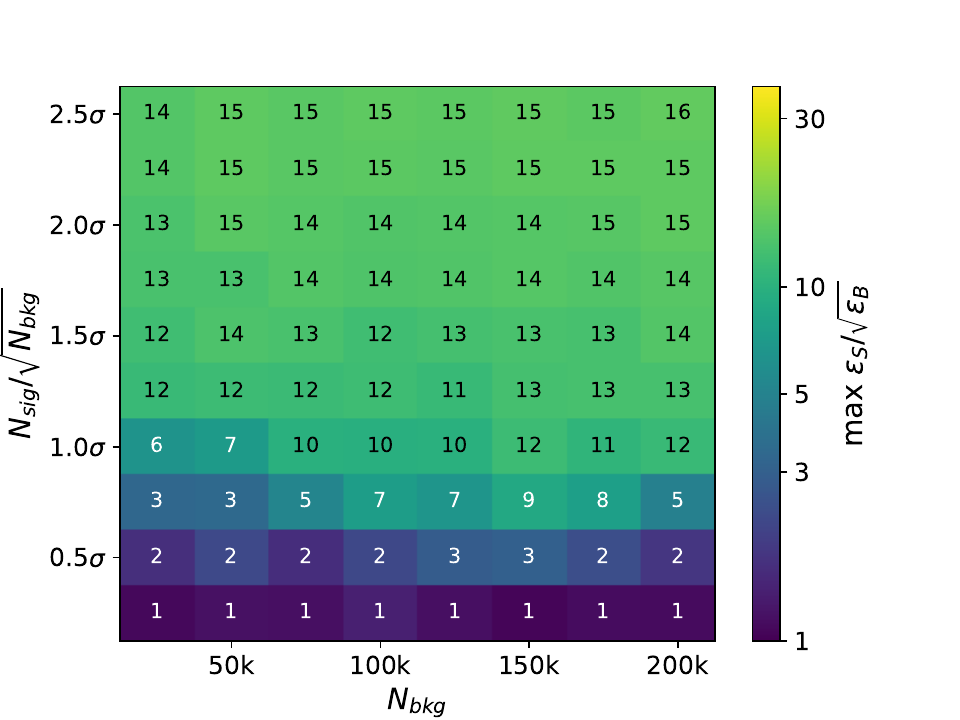}
        \includegraphics[width=0.495\textwidth,trim=0 0 20 30, clip=]{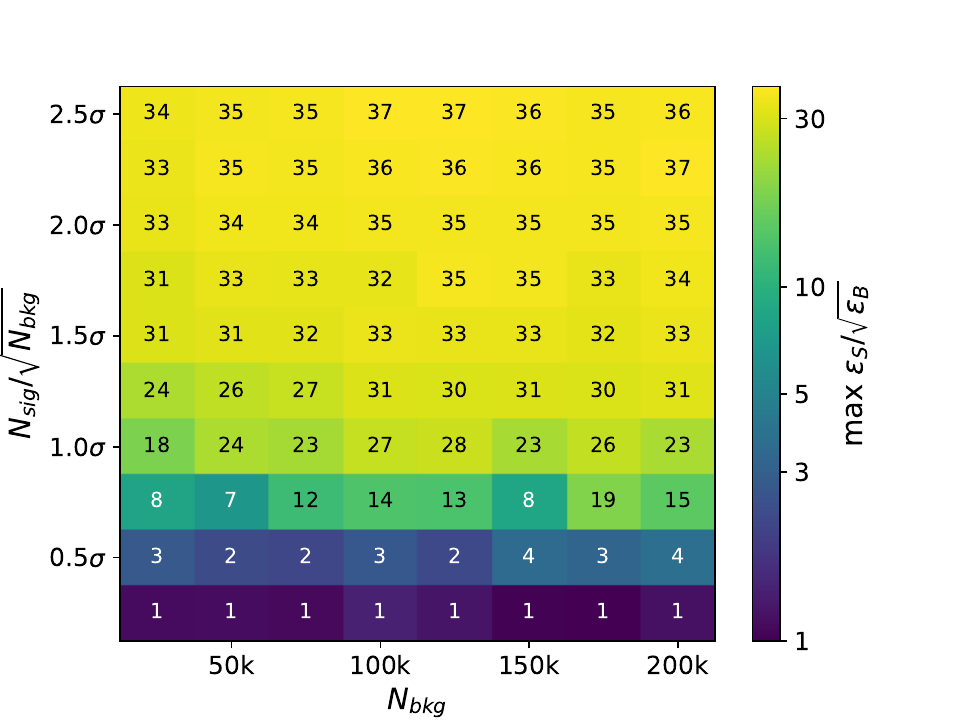}
        \caption{The maximum SIC values as a function of the number of signal events $N_{sig}$ in the SR and the number of background events $N_{bkg}$ in the SR and the background template. The BDT IAD performance is shown for the baseline feature set (upper panel) and the more inclusive extended set 3 (lower panel).}
        \label{fig:2D_scans}
\end{figure}
Figure~\ref{fig:2D_scans} shows the maximum improvement in significance for the IAD BDT as a function of the number of signal and background events in the signal region. In contrast to the previous results, for which the background template is larger than the signal region (see Section~\ref{sec:dataset}) in accordance with the studies in Ref.~\cite{Hallin:2021wme}, in Fig.~\ref{fig:2D_scans} the size of the background template is equal to the number of background events $N_{bkg}$ in the SR such that a larger range of $N_{bkg}$ may be scanned with the existing data. The top panel shows the results for the baseline setup, and the bottom panel shows the results for the inclusive set of all subjettiness features (extended set 3). In both panels it can be seen that the maximum significance improvement for a given signal significance $S/\sqrt{B}$ is largely independent of the absolute number of background events. Hence, for larger datasets, the same significance improvement is achieved with a smaller absolute signal fraction. Comparing the top and bottom panels, the dramatic increase in performance of the BDT classifier is evident, with SIC values increasing from 15 for the small feature set to about 40 for the extended feature set. Even more importantly, the minimal significance in the original dataset for a discovery with high significance is decreased quite a bit.

\subsection{Three-prong signal}
\label{subsec:3-prong}

\begin{figure*}[ht]
        \centering
        \includegraphics[width=0.495\textwidth,trim=0 0 0 30, clip=]{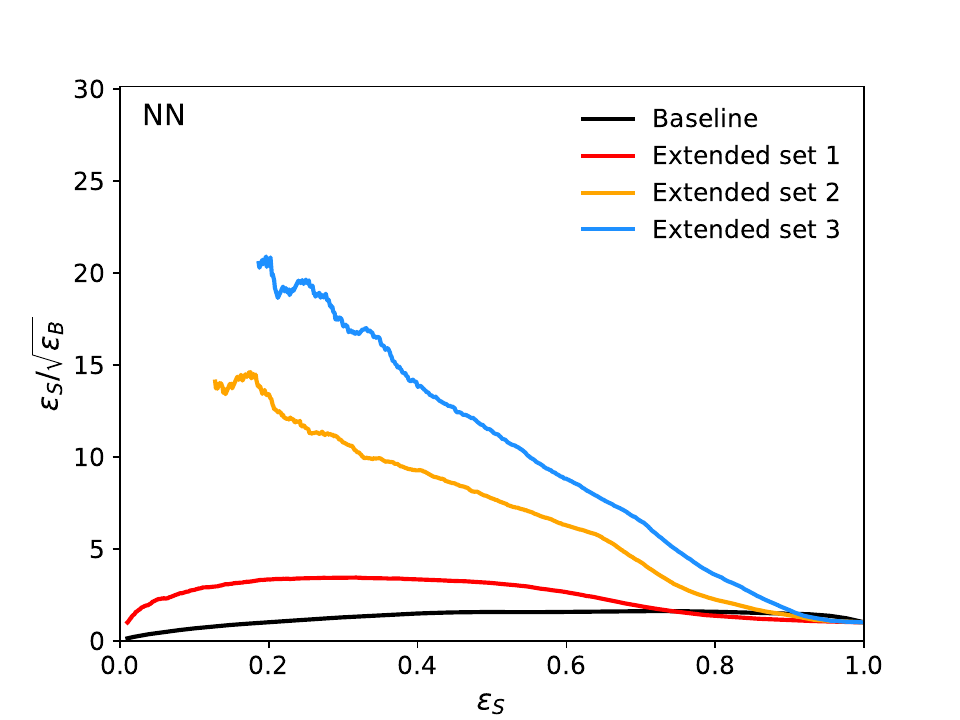}
        \includegraphics[width=0.495\textwidth,trim=0 0 0 30, clip=]{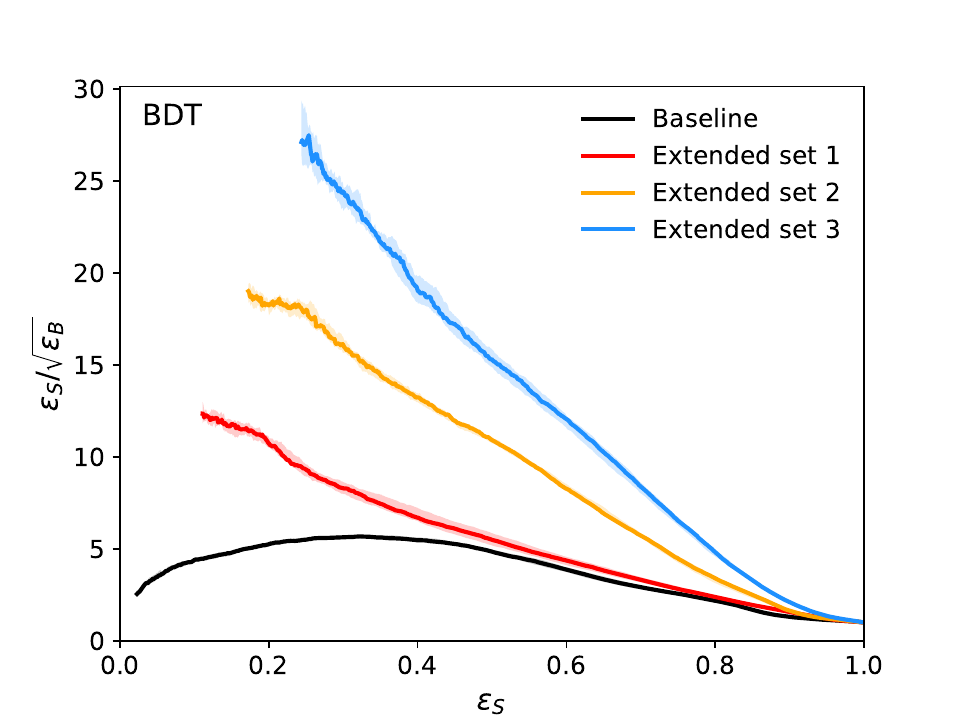}
        \caption{Same as Fig.~\ref{fig:sic-extended_sets} but for a signal model with three-prong jets.}   
        \label{fig:three-prong}
\end{figure*}

So far, we have focused on the LHCO R\&D signal consisting of 2-pronged jets. Here we will compare the performance of NN vs.\ BDT based classifiers across our different feature sets for a different signal (also from the LHCO R\&D dataset) consisting of 3-pronged jets. Clearly, an anomaly detector that claims to be both high performing and model agnostic should be able to find both signals with high sensitivity.

Figure~\ref{fig:three-prong} summarizes our results on the 3-prong signal. 
We see that with the baseline feature set, the NN is essentially no better than random on the 3-prong signal. The $\tau_{21}$ features of the baseline set offer very little discrimination power for the 3-prong signal (i.e.\ are essentially uninformative noise features), so they are evidently degrading the performance of the NN. Meanwhile, the BDT is able to achieve a modest significance improvement, presumably by leveraging the mass features and not being confused by the uninformative $\tau_{21}$ features.

The comparison between NN and BDT is even more dramatic with extended set 1. The NN classifier again shows its noise sensitivity:  while extended set 1 includes the separating $\tau_{32}$ subjettiness ratios for each jet, it also includes higher subjettiness ratios. These are essentially uninformative noise features for the 3-prong signal, and thereby prevent better classification with the NN. However, once again, the BDT does not suffer from this problem, and achieves a significant performance gain relative to the baseline features. 

Using the subjettinesses in extended sets 2 and 3 instead of the ratios yields significantly better NN and BDT classifiers, with the BDT again achieving the best overall performance. Comparing with Fig.~\ref{fig:sic-extended_sets}, we see that NN and BDT classifiers trained on extended sets 2 and 3 perform well as anomaly detectors on both 2-prong and 3-prong signals and can reasonably be called ``model agnostic".

\section{Conclusion}
\label{sec:conclusions}

In this paper we have investigated the role of different machine learning methods (deep learning vs.\ boosted decision trees) on classification performance in weakly supervised anomaly detection. Normally, with the large amount of data and high-dimensional, structured feature sets we have in high energy physics, deep learning wins over BDTs hands down, and this has been widely presumed in the literature since the deep learning revolution came to our field. However, weakly supervised anomaly detection presents special challenges, for which, surprisingly, deep learning may not be ideally suited. Although the amount of data is nominally large, in this anomaly detection context one is trying to detect minute differences between data and background---the amount of signal is small, and so in a sense the effective dataset size is small. Furthermore, in the anomaly detection problem,  one wants to be model agnostic, so one wants a broad selection of features, and for a given signal model maybe only a small subset of them are actually informative (able to discriminate between signal and background). The rest of the features are uninformative noise.  It is well known and well studied in the ML literature that BDTs can outperform deep learning in these settings---tabular data, small to medium dataset sizes, and in the presence of noisy, uninformative features, see e.g.\ \cite{grinsztajn2022treebased}.  

In this work, we have confirmed this in the context of weakly supervised anomaly detection, applied to the setting of the LHC. We have seen that adding by hand a number of uninformative Gaussian noise features to both signal and background completely destroys the performance of an NN classifier, but the anomaly detection ability of a BDT classifier remains robust for at least up to $\sim 50$ Gaussian noise features. We then considered the performance of the idealized anomaly detector with the addition of more physics-motivated features (additional n-subjettinesses $\tau_{n=3,\dots,9}$, although other families of features such as e.g.\ energy flow polynomials~\cite{Komiske:2018cqr} would be possible as well). The NN is unstable with respect to additional features and its performance can deteriorate or improve depending on which features are added. Meanwhile the BDT's performance is more stable and consistently improves as more features are added (up to a certain point---see Fig.~\ref{fig:sic-Gaussian_features}). Finally, we have shown that with more n-subjettiness features, both NNs and BDTs can in principle detect both 2-prong and 3-prong signals, better realizing the original promise of model-agnostic idealized anomaly detection. 

One issue we encountered is the lack of reliable schemes for choosing anomaly detection methods.
We are interested in such model-agnostic metrics for model selection; this is work in progress. 
Also, the issue of $m_{JJ}$ sculpting was left for future work---this important issues should be studied in more detail and something like \lacathode~\cite{Hallin:2022eoq} may be needed to mitigate the sculpting problems. 

Furthermore, this work was focused on the idealized situation where the background templates are assumed to be perfectly modeled. This allowed us to fairly and evenly compare the performance of BDTs and NNs across different feature sets. However, in any realistic scenario, the background template must be derived from data (e.g.\ fully data-driven side-band interpolation as in \cite{Nachman:2020lpy,Hallin:2021wme,Raine:2022hht,Hallin:2022eoq} or simulation-assisted interpolation as in \cite{Andreassen:2020nkr,1815227,Golling:2022nkl}). An essential next step in this study of weakly supervised, model-agnostic anomaly detection is to consider the performance of BDT and NN classifiers using various feature sets that are realistically derived from data.

The prospects for this study are very bright. These techniques (BDT classifiers) can be easily applied to real LHC data, searches based on CWoLa hunting by ATLAS already exist~\cite{ATLAS:2020iwa}, and their physical performance should be significantly improved using BDTs and larger feature sets.  Although it was not the primary objective of this work, we also observe that BDTs are much more efficient in their use of computational resources than NNs---they can be trained and evaluated in a fraction of the time. This could be enormously beneficial to real searches (such as \cite{ATLAS:2020iwa}) that use weak supervision, 
where a very large number of classifiers typically need to be trained and evaluated, especially for the estimation of systematic uncertainties.\\

{\bf Note added:} While this work was being completed, we learned of the work of \cite{otherBDT} who also studied the applications of BDTs to weakly supervised anomaly detection at the LHC.

\section*{Acknowledgements}
TF, MH, MK, and AM would like to thank Thea Klaeboe Aarrestad for sparking our interest in boosted decision trees. TF is supported by the Deutsche Forschungsgemeinschaft (DFG, German Research Foundation) under grant 400140256 - GRK 2497: The physics of the heaviest particles at the Large Hadron Collider. The research of MH, MK and AM is supported by the DFG under grant 396021762 - TRR 257: Particle Physics Phenomenology after the Higgs Discovery. GK, MS, and TQ acknowledge support by the DFG under Germany’s Excellence Strategy 390833306 – EXC 2121: Quantum Universe. DS is supported by DOE grant DOE-SC0010008.
Computations were performed with computing resources granted by RWTH Aachen University under project rwth0934.

\section*{Code}
The code for this paper can be found at \href{https://github.com/uhh-pd-ml/treebased_anomaly_detection}{\texttt{https://github.com/uhh-pd-ml/treebased\_anomaly\_ detection}}.

\appendix

\section{Comparison of different BDT architectures}\label{app:bdt_hyper}

Here, we investigate the performance of other tree-based algorithms in addition to the HGBDT used in the main text. The considered models are:

\begin{itemize}

\item HGBDT (Histogrammed Gradient BDT): A histogrammed version of gradient boosted decision trees. Gradient boosting trains subsequent models on the residuals of the previous ensemble state. The \texttt{HistGradientBoostingClassifier} implementation in \texttt{scikit-learn}~\cite{Pedregosa:2011sk} is used.

\item Random forest: Multiple independent decision trees are built based on subsampling and use a majority vote for classification. The \texttt{RandomForestClassifier} implementation in \texttt{scikit-learn}~\cite{Pedregosa:2011sk} is used.

\item Adaboost: A boosting algorithm where decision trees are iteratively added to improve classification results. Weights are introduced for the trees as well as for the samples to focus on misclassified samples. The \texttt{AdaBoostClassifier} implementation in \texttt{scikit-learn}~\cite{Pedregosa:2011sk} is used.

\item ROOT TMVA BDT: A BDT implementation that is frequently used in Particle Physics applications. Here, we us version 6.28.4~\cite{Brun:1997}. Using the default settings, this implementation is also based on Adaboost.

\end{itemize}

\begin{figure*}[ht]
        \centering
        \includegraphics[width=0.495\textwidth,trim=0 0 0 30, clip=]{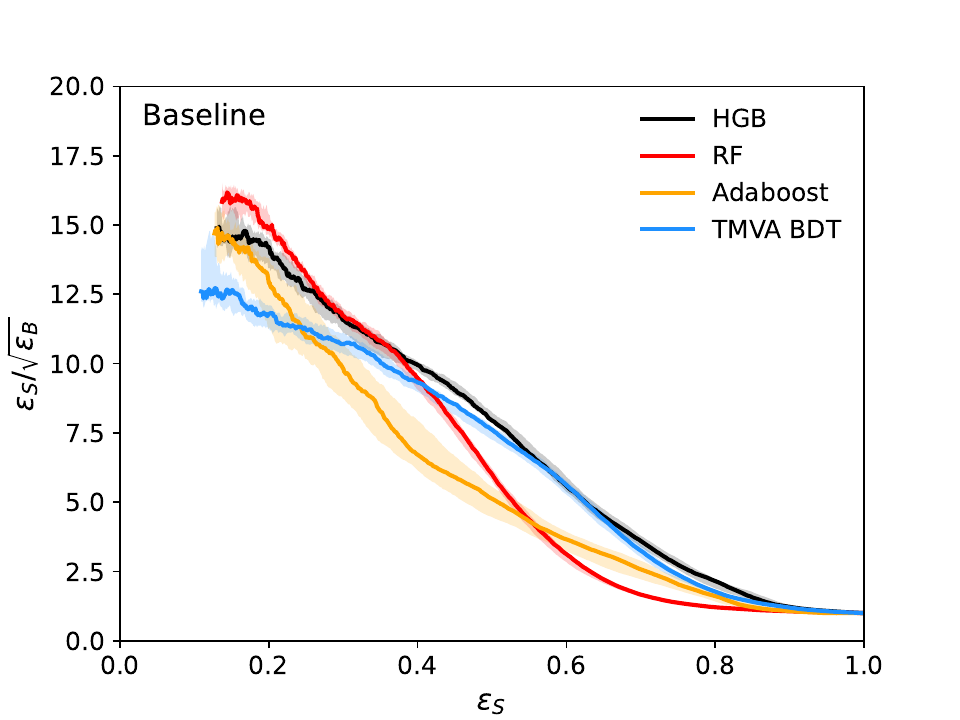}
        \includegraphics[width=0.495\textwidth,trim=0 0 0 30, clip=]{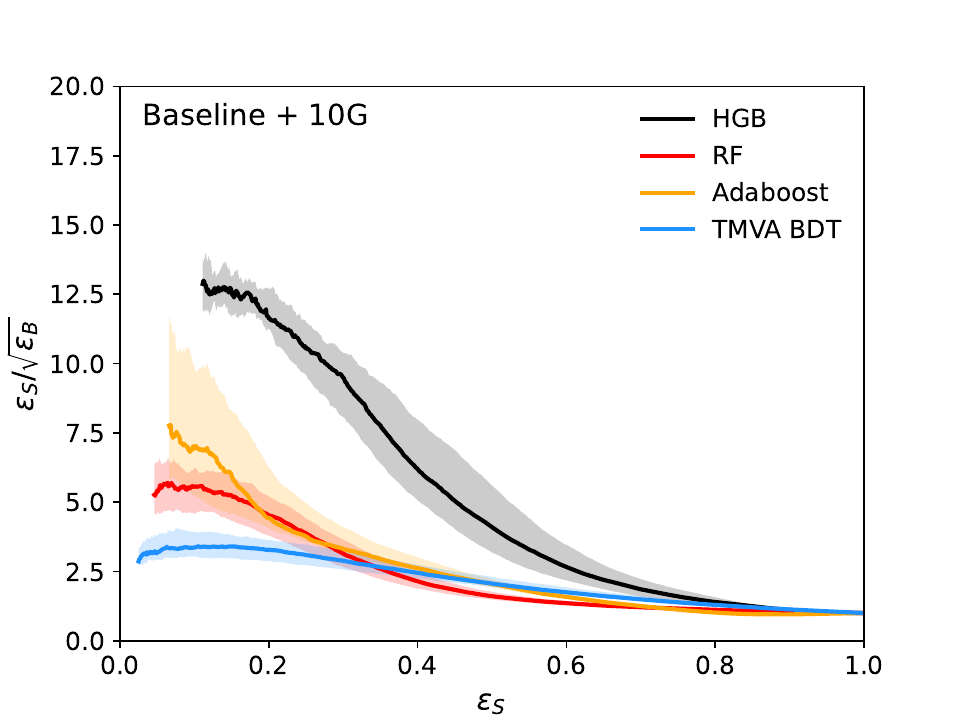}
        \caption{Comparison of different tree-based algorithms for the baseline setup (left panel) as well as for the baseline setup with additional 10 Gaussian noise features (right panel). We show results for the HGBDT used in the main text, a random forest (RF), an Adaboost classifier as well as the BDT classifier implementation in ROOT TMVA \cite{Brun:1997}.}
        \label{fig:algo-compare}
\end{figure*}

For the weakly supervised classification both with and without additional noise features, the results are summarized in Fig.~\ref{fig:algo-compare}. 
As in the main text ten classifiers are used for creating the SIC curve and the corresponding error bands. However, instead of using ensembling with $N=50$, we here use $N=10$ because not all models are fast to train. For the ROOT TMVA model, we used the default settings. The optimal parameters of the other models were tuned based on the validation loss of the weak classification task without noise using the optuna software package \cite{Akiba:2019}. For the GBDT, we find that the default parameters described in Section~\ref{sec:bdts} are close to optimal such that we do not use any optimization in the main text.

Without noise, all tree-based algorithms show similar performance, in particular for low signal efficiencies. When adding 10 Gaussian noise features, however, a clear performance gap can be seen: While the Adaboost, random forest and ROOT TMVA models drop significantly, the histogrammed GBDT retains much of its original performance. Due to the superiority of the histogrammed GBDT in these noise studies, we use it throughout this work. These results are compatible with what is generally considered the state of the art in DT based models \cite{Carlens:2023state}.

\section{Uninformative features, rotational invariance, and tabular data}\label{app:uninformative}

\begin{figure*}[ht]
        \centering
        \includegraphics[width=0.495\textwidth,trim=0 0 0 30, clip=]{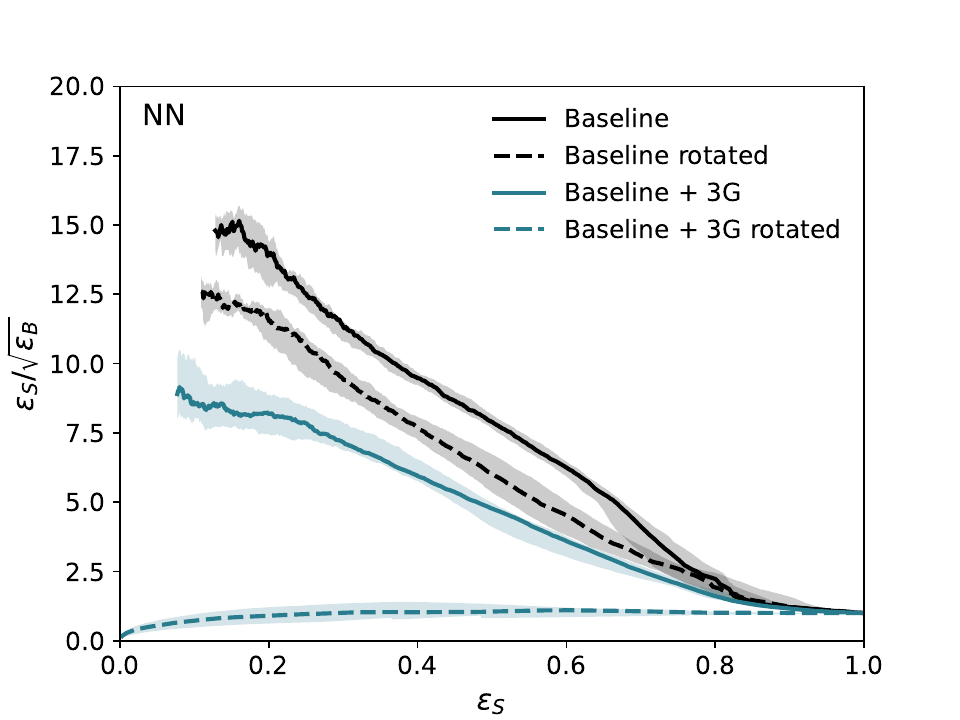}
        \includegraphics[width=0.495\textwidth,trim=0 0 0 30, clip=]{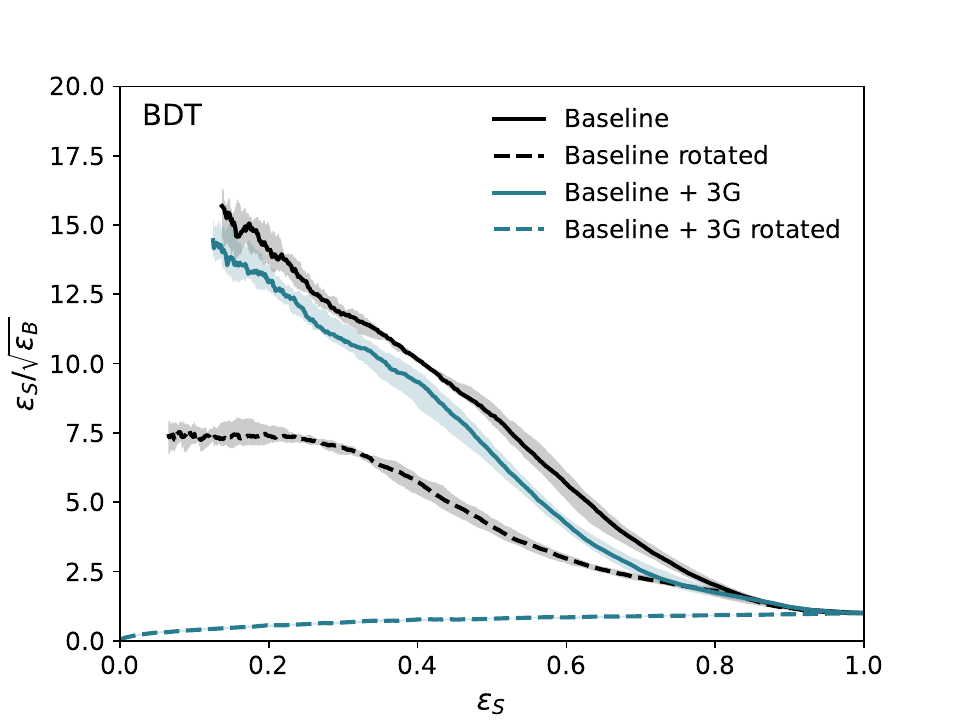}
        \caption{The impact of a random rotation on the BDT IAD and NN IAD using the baseline feature set and the baseline features with additional 3 Gaussian noise features. SIC curves are compared with and without a random rotation applied for NN models (left panel) and BDT models (right panel).}
        \label{fig:rotational-invariance}
\end{figure*}

In \cite{grinsztajn2022treebased}, a key argument for decision tree (DT) based models outperforming deep learning on tabular data is their robustness to uninformative features. This robustness stems from the fact that tabular data is not invariant under rotation, and therefore an algorithm with the same property should be used to learn from it. Since NNs are rotationally invariant \cite{Ng:10.1145/1015330.1015435}, they must learn the ideal feature orientation in an increasingly high-dimensional input space and then identify the most informative features. The information in the original orientation of the data, which in our case is physics based, is lost. On the other hand, DT-based models are not rotationally invariant and therefore only operate on the correct orientation with respect to the physics of the event.

Since the data used in this work is tabular, we expect the same behavior in the weakly supervised setting: While the performance of BDT-based models should break down significantly under a random rotation of the data, NN-based models should still show a similar performance. However, if no rotation is applied, BDT-based models should outperform NNs, and the performance gap is expected to increase significantly as more uninformative features are used.

Figure~\ref{fig:rotational-invariance} shows the performance of both BDT and NN with and without a random rotation applied to the input features. As in Appendix~\ref{app:bdt_hyper}, the medians and error bands are based on ten classifiers using ensembling with $N=10$. We observe the expected drop in performance for a BDT trained on rotated features, where the maximum SIC value on the baseline features is halved. When three noise features are added, the BDT is unable to maintain better than random performance. A slight drop in SIC of about 2 is also observed for the NN on the rotated features. With the addition of noise features, the drop is slightly larger, but the NN still remains better than random.

\section{Ensembling}\label{app:ensemble}
\begin{figure*}[ht]
        \centering
        \includegraphics[width=0.495\textwidth,trim=0 0 0 30, clip=]{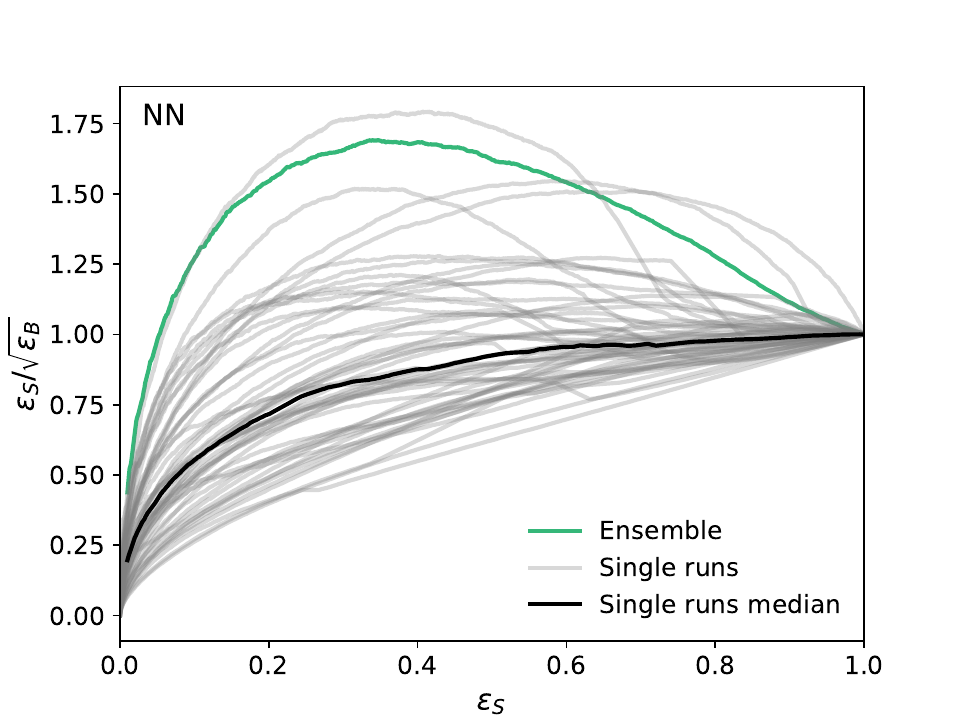}
        \includegraphics[width=0.495\textwidth,trim=0 0 0 30, clip=]{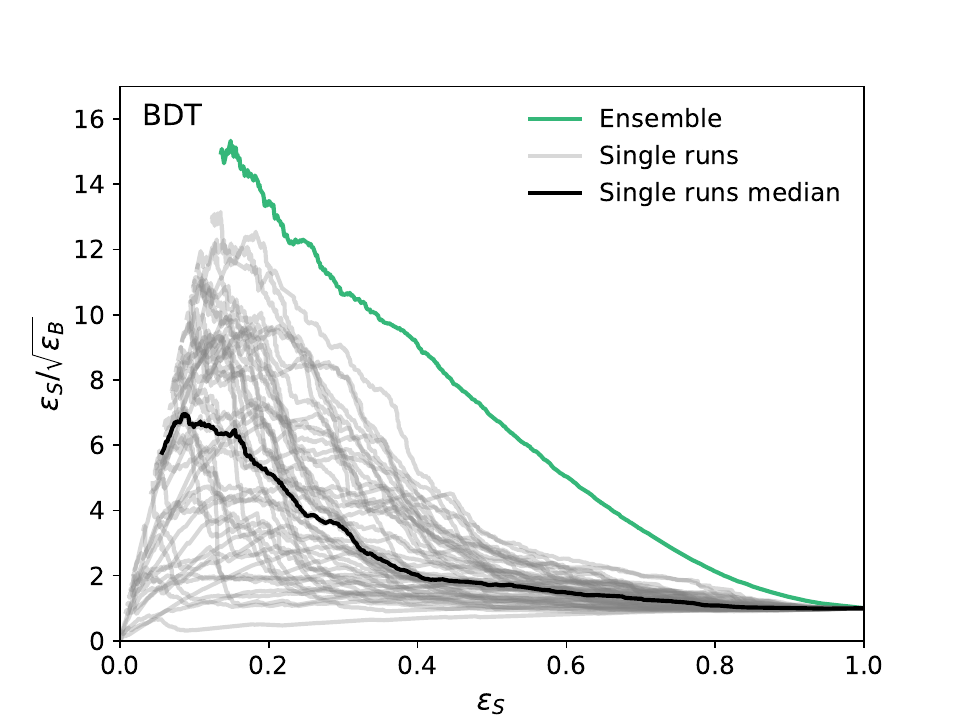}        \caption{The impact of ensembling on the baseline setup with four features and additional 10 Gaussian noise features for both NN (left panel) and BDT (right panel). The gray lines show individual SICs of single BDTs/NNs. Additionally, the median performance of the individual BDTs/NNs and the performance of the ensemble classifiers are shown. Note that for the BDT the ensemble classifier is one of the classifiers used to calculate the median shown in Fig.~\ref{fig:sic-Gaussian_features}.}
        \label{fig:sic-ensembling-10G}
\end{figure*}

In general, the use of ensembles of different classifiers is a well-established technique in many machine learning applications. A single BDT is already a form of ensemble learning, as boosting combines different decision trees. Often, different ML models are combined in an ensemble to exploit their different strengths. Typically, the ensemble outperforms each of the individual models. This can be seen for example in the comparison of the top taggers in \cite{Kasieczka:2019dbj}. 

In weakly supervised anomaly detection, the goal is to lower the threshold for detecting a small fraction of signal in an overwhelming background. Around this threshold, the performance of classifiers becomes unstable. The statistical fluctuation in the assignment of signal events to the training or validation set can be large enough to have a significant impact on the quality of the classification. In addition, subsampling is used in our BDT implementation, resulting in variability in performance. For NNs, the random initialization of the network has a similar effect. The large variance in individual BDT or NN predictions close to the signal fraction threshold requires the use of an ensemble classifier as described in Sections~\ref{sec:bdts} and \ref{sec:nn}.  

For the BDT classifier with an ensembling of $N=50$, the performance is significantly increased even with respect to the best individual training, as shown in Fig.~\ref{fig:sic-ensembling-10G} for the default baseline setup with 10 additional Gaussian noise features. The variance of the ensemble classifier is quite small as shown in Fig.~\ref{fig:sic-Gaussian_features}. For the NN, an increase in the performance of the ensemble classifier is also observed. However, since no single run achieves a max SIC value significantly above 2, the NN ensemble cannot match the performance of the BDT ensemble.

In addition, considering the very different computational costs of training BDTs and NNs, the BDT ensemble not only performs better, but is also much more cost effective.

\bibliography{HEPML, other}

\begin{thebibliography}{39}%
\makeatletter
\providecommand \@ifxundefined [1]{%
 \@ifx{#1\undefined}
}%
\providecommand \@ifnum [1]{%
 \ifnum #1\expandafter \@firstoftwo
 \else \expandafter \@secondoftwo
 \fi
}%
\providecommand \@ifx [1]{%
 \ifx #1\expandafter \@firstoftwo
 \else \expandafter \@secondoftwo
 \fi
}%
\providecommand \natexlab [1]{#1}%
\providecommand \enquote  [1]{``#1''}%
\providecommand \bibnamefont  [1]{#1}%
\providecommand \bibfnamefont [1]{#1}%
\providecommand \citenamefont [1]{#1}%
\providecommand \href@noop [0]{\@secondoftwo}%
\providecommand \href [0]{\begingroup \@sanitize@url \@href}%
\providecommand \@href[1]{\@@startlink{#1}\@@href}%
\providecommand \@@href[1]{\endgroup#1\@@endlink}%
\providecommand \@sanitize@url [0]{\catcode `\\12\catcode `\$12\catcode `\&12\catcode `\#12\catcode `\^12\catcode `\_12\catcode `\%12\relax}%
\providecommand \@@startlink[1]{}%
\providecommand \@@endlink[0]{}%
\providecommand \url  [0]{\begingroup\@sanitize@url \@url }%
\providecommand \@url [1]{\endgroup\@href {#1}{\urlprefix }}%
\providecommand \urlprefix  [0]{URL }%
\providecommand \Eprint [0]{\href }%
\providecommand \doibase [0]{http://dx.doi.org/}%
\providecommand \selectlanguage [0]{\@gobble}%
\providecommand \bibinfo  [0]{\@secondoftwo}%
\providecommand \bibfield  [0]{\@secondoftwo}%
\providecommand \translation [1]{[#1]}%
\providecommand \BibitemOpen [0]{}%
\providecommand \bibitemStop [0]{}%
\providecommand \bibitemNoStop [0]{.\EOS\space}%
\providecommand \EOS [0]{\spacefactor3000\relax}%
\providecommand \BibitemShut  [1]{\csname bibitem#1\endcsname}%
\let\auto@bib@innerbib\@empty
\bibitem [{\citenamefont {Kasieczka}\ \emph {et~al.}()\citenamefont {Kasieczka} \emph {et~al.}}]{Kasieczka:2021xcg}%
  \BibitemOpen
  \bibfield  {author} {\bibinfo {author} {\bibfnamefont {G.}~\bibnamefont {Kasieczka}} \emph {et~al.},\ }\href@noop {} {\ }\Eprint {http://arxiv.org/abs/2101.08320} {arXiv:2101.08320 [hep-ph]} \BibitemShut {NoStop}%
\bibitem [{\citenamefont {Aarrestad}\ \emph {et~al.}()\citenamefont {Aarrestad} \emph {et~al.}}]{Aarrestad:2021oeb}%
  \BibitemOpen
  \bibfield  {author} {\bibinfo {author} {\bibfnamefont {T.}~\bibnamefont {Aarrestad}} \emph {et~al.},\ }\href@noop {} {\ }\Eprint {http://arxiv.org/abs/2105.14027} {arXiv:2105.14027 [hep-ph]} \BibitemShut {NoStop}%
\bibitem [{\citenamefont {Metodiev}\ \emph {et~al.}(2017)\citenamefont {Metodiev}, \citenamefont {Nachman},\ and\ \citenamefont {Thaler}}]{Metodiev:2017vrx}%
  \BibitemOpen
  \bibfield  {author} {\bibinfo {author} {\bibfnamefont {E.~M.}\ \bibnamefont {Metodiev}}, \bibinfo {author} {\bibfnamefont {B.}~\bibnamefont {Nachman}}, \ and\ \bibinfo {author} {\bibfnamefont {J.}~\bibnamefont {Thaler}},\ }\href {\doibase 10.1007/JHEP10(2017)174} {\bibfield  {journal} {\bibinfo  {journal} {JHEP}\ }\textbf {\bibinfo {volume} {10}},\ \bibinfo {pages} {174} (\bibinfo {year} {2017})},\ \Eprint {http://arxiv.org/abs/1708.02949} {arXiv:1708.02949 [hep-ph]} \BibitemShut {NoStop}%
\bibitem [{\citenamefont {Collins}\ \emph {et~al.}(2018)\citenamefont {Collins}, \citenamefont {Howe},\ and\ \citenamefont {Nachman}}]{Collins:2018epr}%
  \BibitemOpen
  \bibfield  {author} {\bibinfo {author} {\bibfnamefont {J.~H.}\ \bibnamefont {Collins}}, \bibinfo {author} {\bibfnamefont {K.}~\bibnamefont {Howe}}, \ and\ \bibinfo {author} {\bibfnamefont {B.}~\bibnamefont {Nachman}},\ }\href {\doibase 10.1103/PhysRevLett.121.241803} {\bibfield  {journal} {\bibinfo  {journal} {Phys. Rev. Lett.}\ }\textbf {\bibinfo {volume} {121}},\ \bibinfo {pages} {241803} (\bibinfo {year} {2018})},\ \Eprint {http://arxiv.org/abs/1805.02664} {arXiv:1805.02664 [hep-ph]} \BibitemShut {NoStop}%
\bibitem [{\citenamefont {Collins}\ \emph {et~al.}(2019)\citenamefont {Collins}, \citenamefont {Howe},\ and\ \citenamefont {Nachman}}]{Collins:2019jip}%
  \BibitemOpen
  \bibfield  {author} {\bibinfo {author} {\bibfnamefont {J.~H.}\ \bibnamefont {Collins}}, \bibinfo {author} {\bibfnamefont {K.}~\bibnamefont {Howe}}, \ and\ \bibinfo {author} {\bibfnamefont {B.}~\bibnamefont {Nachman}},\ }\href {\doibase 10.1103/PhysRevD.99.014038} {\bibfield  {journal} {\bibinfo  {journal} {Phys. Rev.}\ }\textbf {\bibinfo {volume} {D99}},\ \bibinfo {pages} {014038} (\bibinfo {year} {2019})},\ \Eprint {http://arxiv.org/abs/1902.02634} {arXiv:1902.02634 [hep-ph]} \BibitemShut {NoStop}%
\bibitem [{\citenamefont {Nachman}\ and\ \citenamefont {Shih}(2020)}]{Nachman:2020lpy}%
  \BibitemOpen
  \bibfield  {author} {\bibinfo {author} {\bibfnamefont {B.}~\bibnamefont {Nachman}}\ and\ \bibinfo {author} {\bibfnamefont {D.}~\bibnamefont {Shih}},\ }\href {\doibase 10.1103/PhysRevD.101.075042} {\bibfield  {journal} {\bibinfo  {journal} {Phys. Rev. D}\ }\textbf {\bibinfo {volume} {101}},\ \bibinfo {pages} {075042} (\bibinfo {year} {2020})},\ \Eprint {http://arxiv.org/abs/2001.04990} {arXiv:2001.04990 [hep-ph]} \BibitemShut {NoStop}%
\bibitem [{\citenamefont {Andreassen}\ \emph {et~al.}(2020)\citenamefont {Andreassen}, \citenamefont {Nachman},\ and\ \citenamefont {Shih}}]{Andreassen:2020nkr}%
  \BibitemOpen
  \bibfield  {author} {\bibinfo {author} {\bibfnamefont {A.}~\bibnamefont {Andreassen}}, \bibinfo {author} {\bibfnamefont {B.}~\bibnamefont {Nachman}}, \ and\ \bibinfo {author} {\bibfnamefont {D.}~\bibnamefont {Shih}},\ }\href {\doibase 10.1103/PhysRevD.101.095004} {\bibfield  {journal} {\bibinfo  {journal} {Phys. Rev. D}\ }\textbf {\bibinfo {volume} {101}},\ \bibinfo {pages} {095004} (\bibinfo {year} {2020})},\ \Eprint {http://arxiv.org/abs/2001.05001} {arXiv:2001.05001 [hep-ph]} \BibitemShut {NoStop}%
\bibitem [{\citenamefont {Benkendorfer}\ \emph {et~al.}(2021)\citenamefont {Benkendorfer}, \citenamefont {Pottier},\ and\ \citenamefont {Nachman}}]{1815227}%
  \BibitemOpen
  \bibfield  {author} {\bibinfo {author} {\bibfnamefont {K.}~\bibnamefont {Benkendorfer}}, \bibinfo {author} {\bibfnamefont {L.~L.}\ \bibnamefont {Pottier}}, \ and\ \bibinfo {author} {\bibfnamefont {B.}~\bibnamefont {Nachman}},\ }\href {\doibase 10.1103/PhysRevD.104.035003} {\bibfield  {journal} {\bibinfo  {journal} {Phys. Rev. D}\ }\textbf {\bibinfo {volume} {104}},\ \bibinfo {pages} {035003} (\bibinfo {year} {2021})},\ \Eprint {http://arxiv.org/abs/2009.02205} {arXiv:2009.02205 [hep-ph]} \BibitemShut {NoStop}%
\bibitem [{\citenamefont {Hallin}\ \emph {et~al.}(2022)\citenamefont {Hallin}, \citenamefont {Isaacson}, \citenamefont {Kasieczka}, \citenamefont {Krause}, \citenamefont {Nachman}, \citenamefont {Quadfasel}, \citenamefont {Schlaffer}, \citenamefont {Shih},\ and\ \citenamefont {Sommerhalder}}]{Hallin:2021wme}%
  \BibitemOpen
  \bibfield  {author} {\bibinfo {author} {\bibfnamefont {A.}~\bibnamefont {Hallin}}, \bibinfo {author} {\bibfnamefont {J.}~\bibnamefont {Isaacson}}, \bibinfo {author} {\bibfnamefont {G.}~\bibnamefont {Kasieczka}}, \bibinfo {author} {\bibfnamefont {C.}~\bibnamefont {Krause}}, \bibinfo {author} {\bibfnamefont {B.}~\bibnamefont {Nachman}}, \bibinfo {author} {\bibfnamefont {T.}~\bibnamefont {Quadfasel}}, \bibinfo {author} {\bibfnamefont {M.}~\bibnamefont {Schlaffer}}, \bibinfo {author} {\bibfnamefont {D.}~\bibnamefont {Shih}}, \ and\ \bibinfo {author} {\bibfnamefont {M.}~\bibnamefont {Sommerhalder}},\ }\href {\doibase 10.1103/PhysRevD.106.055006} {\bibfield  {journal} {\bibinfo  {journal} {Phys. Rev. D}\ }\textbf {\bibinfo {volume} {106}},\ \bibinfo {pages} {055006} (\bibinfo {year} {2022})},\ \Eprint {http://arxiv.org/abs/2109.00546} {arXiv:2109.00546 [hep-ph]} \BibitemShut {NoStop}%
\bibitem [{\citenamefont {Raine}\ \emph {et~al.}(2023)\citenamefont {Raine}, \citenamefont {Klein}, \citenamefont {Sengupta},\ and\ \citenamefont {Golling}}]{Raine:2022hht}%
  \BibitemOpen
  \bibfield  {author} {\bibinfo {author} {\bibfnamefont {J.~A.}\ \bibnamefont {Raine}}, \bibinfo {author} {\bibfnamefont {S.}~\bibnamefont {Klein}}, \bibinfo {author} {\bibfnamefont {D.}~\bibnamefont {Sengupta}}, \ and\ \bibinfo {author} {\bibfnamefont {T.}~\bibnamefont {Golling}},\ }\href {\doibase 10.3389/fdata.2023.899345} {\bibfield  {journal} {\bibinfo  {journal} {Front. Big Data}\ }\textbf {\bibinfo {volume} {6}},\ \bibinfo {pages} {899345} (\bibinfo {year} {2023})},\ \Eprint {http://arxiv.org/abs/2203.09470} {arXiv:2203.09470 [hep-ph]} \BibitemShut {NoStop}%
\bibitem [{\citenamefont {Hallin}\ \emph {et~al.}(2023)\citenamefont {Hallin}, \citenamefont {Kasieczka}, \citenamefont {Quadfasel}, \citenamefont {Shih},\ and\ \citenamefont {Sommerhalder}}]{Hallin:2022eoq}%
  \BibitemOpen
  \bibfield  {author} {\bibinfo {author} {\bibfnamefont {A.}~\bibnamefont {Hallin}}, \bibinfo {author} {\bibfnamefont {G.}~\bibnamefont {Kasieczka}}, \bibinfo {author} {\bibfnamefont {T.}~\bibnamefont {Quadfasel}}, \bibinfo {author} {\bibfnamefont {D.}~\bibnamefont {Shih}}, \ and\ \bibinfo {author} {\bibfnamefont {M.}~\bibnamefont {Sommerhalder}},\ }\href {\doibase 10.1103/PhysRevD.107.114012} {\bibfield  {journal} {\bibinfo  {journal} {Phys. Rev. D}\ }\textbf {\bibinfo {volume} {107}},\ \bibinfo {pages} {114012} (\bibinfo {year} {2023})},\ \Eprint {http://arxiv.org/abs/2210.14924} {arXiv:2210.14924 [hep-ph]} \BibitemShut {NoStop}%
\bibitem [{\citenamefont {Golling}\ \emph {et~al.}(2023)\citenamefont {Golling}, \citenamefont {Klein}, \citenamefont {Mastandrea},\ and\ \citenamefont {Nachman}}]{Golling:2022nkl}%
  \BibitemOpen
  \bibfield  {author} {\bibinfo {author} {\bibfnamefont {T.}~\bibnamefont {Golling}}, \bibinfo {author} {\bibfnamefont {S.}~\bibnamefont {Klein}}, \bibinfo {author} {\bibfnamefont {R.}~\bibnamefont {Mastandrea}}, \ and\ \bibinfo {author} {\bibfnamefont {B.}~\bibnamefont {Nachman}},\ }\href {\doibase 10.1103/PhysRevD.107.096025} {\bibfield  {journal} {\bibinfo  {journal} {Phys. Rev. D}\ }\textbf {\bibinfo {volume} {107}},\ \bibinfo {pages} {096025} (\bibinfo {year} {2023})},\ \Eprint {http://arxiv.org/abs/2212.11285} {arXiv:2212.11285 [hep-ph]} \BibitemShut {NoStop}%
\bibitem [{\citenamefont {Golling}\ \emph {et~al.}()\citenamefont {Golling}, \citenamefont {Kasieczka}, \citenamefont {Krause}, \citenamefont {Mastandrea}, \citenamefont {Nachman}, \citenamefont {Raine}, \citenamefont {Sengupta}, \citenamefont {Shih},\ and\ \citenamefont {Sommerhalder}}]{Golling:2023yjq}%
  \BibitemOpen
  \bibfield  {author} {\bibinfo {author} {\bibfnamefont {T.}~\bibnamefont {Golling}}, \bibinfo {author} {\bibfnamefont {G.}~\bibnamefont {Kasieczka}}, \bibinfo {author} {\bibfnamefont {C.}~\bibnamefont {Krause}}, \bibinfo {author} {\bibfnamefont {R.}~\bibnamefont {Mastandrea}}, \bibinfo {author} {\bibfnamefont {B.}~\bibnamefont {Nachman}}, \bibinfo {author} {\bibfnamefont {J.~A.}\ \bibnamefont {Raine}}, \bibinfo {author} {\bibfnamefont {D.}~\bibnamefont {Sengupta}}, \bibinfo {author} {\bibfnamefont {D.}~\bibnamefont {Shih}}, \ and\ \bibinfo {author} {\bibfnamefont {M.}~\bibnamefont {Sommerhalder}},\ }\href@noop {} {\ }\Eprint {http://arxiv.org/abs/2307.11157} {arXiv:2307.11157 [hep-ph]} \BibitemShut {NoStop}%
\bibitem [{\citenamefont {Grinsztajn}\ \emph {et~al.}()\citenamefont {Grinsztajn}, \citenamefont {Oyallon},\ and\ \citenamefont {Varoquaux}}]{grinsztajn2022treebased}%
  \BibitemOpen
  \bibfield  {author} {\bibinfo {author} {\bibfnamefont {L.}~\bibnamefont {Grinsztajn}}, \bibinfo {author} {\bibfnamefont {E.}~\bibnamefont {Oyallon}}, \ and\ \bibinfo {author} {\bibfnamefont {G.}~\bibnamefont {Varoquaux}},\ }\href@noop {} {\ }\Eprint {http://arxiv.org/abs/2207.08815} {arXiv:2207.08815 [cs.LG]} \BibitemShut {NoStop}%
\bibitem [{\citenamefont {Borisov}\ \emph {et~al.}(2022)\citenamefont {Borisov}, \citenamefont {Leemann}, \citenamefont {Seßler}, \citenamefont {Haug}, \citenamefont {Pawelczyk},\ and\ \citenamefont {Kasneci}}]{9998482}%
  \BibitemOpen
  \bibfield  {author} {\bibinfo {author} {\bibfnamefont {V.}~\bibnamefont {Borisov}}, \bibinfo {author} {\bibfnamefont {T.}~\bibnamefont {Leemann}}, \bibinfo {author} {\bibfnamefont {K.}~\bibnamefont {Seßler}}, \bibinfo {author} {\bibfnamefont {J.}~\bibnamefont {Haug}}, \bibinfo {author} {\bibfnamefont {M.}~\bibnamefont {Pawelczyk}}, \ and\ \bibinfo {author} {\bibfnamefont {G.}~\bibnamefont {Kasneci}},\ }\href {\doibase 10.1109/TNNLS.2022.3229161} {\bibfield  {journal} {\bibinfo  {journal} {IEEE Transactions on Neural Networks and Learning Systems}\ ,\ \bibinfo {pages} {1}} (\bibinfo {year} {2022})}\BibitemShut {NoStop}%
\bibitem [{\citenamefont {Neyman}\ and\ \citenamefont {Pearson}(1933)}]{Neyman:1933wgr}%
  \BibitemOpen
  \bibfield  {author} {\bibinfo {author} {\bibfnamefont {J.}~\bibnamefont {Neyman}}\ and\ \bibinfo {author} {\bibfnamefont {E.~S.}\ \bibnamefont {Pearson}},\ }\href {\doibase 10.1098/rsta.1933.0009} {\bibfield  {journal} {\bibinfo  {journal} {Phil. Trans. Roy. Soc. Lond. A}\ }\textbf {\bibinfo {volume} {231}},\ \bibinfo {pages} {289} (\bibinfo {year} {1933})}\BibitemShut {NoStop}%
\bibitem [{\citenamefont {Finke}\ \emph {et~al.}(2022)\citenamefont {Finke}, \citenamefont {Kr\"amer}, \citenamefont {Lipp},\ and\ \citenamefont {M\"uck}}]{Finke:2022lsu}%
  \BibitemOpen
  \bibfield  {author} {\bibinfo {author} {\bibfnamefont {T.}~\bibnamefont {Finke}}, \bibinfo {author} {\bibfnamefont {M.}~\bibnamefont {Kr\"amer}}, \bibinfo {author} {\bibfnamefont {M.}~\bibnamefont {Lipp}}, \ and\ \bibinfo {author} {\bibfnamefont {A.}~\bibnamefont {M\"uck}},\ }\href {\doibase 10.1007/JHEP08(2022)015} {\bibfield  {journal} {\bibinfo  {journal} {JHEP}\ }\textbf {\bibinfo {volume} {08}},\ \bibinfo {pages} {015} (\bibinfo {year} {2022})},\ \Eprint {http://arxiv.org/abs/2204.11889} {arXiv:2204.11889 [hep-ph]} \BibitemShut {NoStop}%
\bibitem [{\citenamefont {Ke}\ \emph {et~al.}(2017)\citenamefont {Ke}, \citenamefont {Meng}, \citenamefont {Finley}, \citenamefont {Wang}, \citenamefont {Chen}, \citenamefont {Ma}, \citenamefont {Ye},\ and\ \citenamefont {Liu}}]{Ke:2017lgbm}%
  \BibitemOpen
  \bibfield  {author} {\bibinfo {author} {\bibfnamefont {G.}~\bibnamefont {Ke}}, \bibinfo {author} {\bibfnamefont {Q.}~\bibnamefont {Meng}}, \bibinfo {author} {\bibfnamefont {T.}~\bibnamefont {Finley}}, \bibinfo {author} {\bibfnamefont {T.}~\bibnamefont {Wang}}, \bibinfo {author} {\bibfnamefont {W.}~\bibnamefont {Chen}}, \bibinfo {author} {\bibfnamefont {W.}~\bibnamefont {Ma}}, \bibinfo {author} {\bibfnamefont {Q.}~\bibnamefont {Ye}}, \ and\ \bibinfo {author} {\bibfnamefont {T.-Y.}\ \bibnamefont {Liu}},\ }\href@noop {} {\bibfield  {journal} {\bibinfo  {journal} {Advances in neural information processing systems}\ }\textbf {\bibinfo {volume} {30}},\ \bibinfo {pages} {3146} (\bibinfo {year} {2017})}\BibitemShut {NoStop}%
\bibitem [{\citenamefont {Carlens}()}]{Carlens:2023state}%
  \BibitemOpen
  \bibfield  {author} {\bibinfo {author} {\bibfnamefont {H.}~\bibnamefont {Carlens}},\ }\href@noop {} {\enquote {\bibinfo {title} {State of competitive machine learning in 2022},}\ }\bibinfo {howpublished} {\url{https://mlcontests.com/state-of-competitive-data-science-2022}}\BibitemShut {NoStop}%
\bibitem [{\citenamefont {Pedregosa}\ \emph {et~al.}(2011)\citenamefont {Pedregosa} \emph {et~al.}}]{Pedregosa:2011sk}%
  \BibitemOpen
  \bibfield  {author} {\bibinfo {author} {\bibfnamefont {F.}~\bibnamefont {Pedregosa}} \emph {et~al.},\ }\href@noop {} {\bibfield  {journal} {\bibinfo  {journal} {Journal of Machine Learning Research}\ }\textbf {\bibinfo {volume} {12}},\ \bibinfo {pages} {2825} (\bibinfo {year} {2011})}\BibitemShut {NoStop}%
\bibitem [{\citenamefont {Abadi}\ \emph {et~al.}(2015)\citenamefont {Abadi} \emph {et~al.}}]{tensorflow2015-whitepaper}%
  \BibitemOpen
  \bibfield  {author} {\bibinfo {author} {\bibfnamefont {M.}~\bibnamefont {Abadi}} \emph {et~al.},\ }\href@noop {} {\enquote {\bibinfo {title} {{TensorFlow}: Large-scale machine learning on heterogeneous systems},}\ } (\bibinfo {year} {2015}),\ \bibinfo {note} {software available from \url{https://www.tensorflow.org/}}\BibitemShut {NoStop}%
\bibitem [{\citenamefont {Chollet}\ \emph {et~al.}(2015)\citenamefont {Chollet} \emph {et~al.}}]{Chollet:2015keras}%
  \BibitemOpen
  \bibfield  {author} {\bibinfo {author} {\bibfnamefont {F.}~\bibnamefont {Chollet}} \emph {et~al.},\ }\href@noop {} {\enquote {\bibinfo {title} {Keras},}\ }\bibinfo {howpublished} {\url{https://keras.io}} (\bibinfo {year} {2015})\BibitemShut {NoStop}%
\bibitem [{\citenamefont {Kingma}\ and\ \citenamefont {Ba}()}]{Kingma:2014ad}%
  \BibitemOpen
  \bibfield  {author} {\bibinfo {author} {\bibfnamefont {D.~P.}\ \bibnamefont {Kingma}}\ and\ \bibinfo {author} {\bibfnamefont {J.}~\bibnamefont {Ba}},\ }\href@noop {} {\ }\Eprint {http://arxiv.org/abs/arXiv:1412.6980} {arXiv:1412.6980} \BibitemShut {NoStop}%
\bibitem [{\citenamefont {Kasieczka}\ \emph {et~al.}(2019)\citenamefont {Kasieczka}, \citenamefont {Nachman},\ and\ \citenamefont {Shih}}]{LHCOdataset}%
  \BibitemOpen
  \bibfield  {author} {\bibinfo {author} {\bibfnamefont {G.}~\bibnamefont {Kasieczka}}, \bibinfo {author} {\bibfnamefont {B.}~\bibnamefont {Nachman}}, \ and\ \bibinfo {author} {\bibfnamefont {D.}~\bibnamefont {Shih}},\ }\href@noop {} {\enquote {\bibinfo {title} {R\&d dataset for lhc olympics 2020 anomaly detection challenge},}\ }\bibinfo {howpublished} {\url{https://zenodo.org/record/6466204}} (\bibinfo {year} {2019})\BibitemShut {NoStop}%
\bibitem [{\citenamefont {Sj{\"o}strand}\ \emph {et~al.}(2006)\citenamefont {Sj{\"o}strand}, \citenamefont {Mrenna},\ and\ \citenamefont {Skands}}]{Sjostrand:2006za}%
  \BibitemOpen
  \bibfield  {author} {\bibinfo {author} {\bibfnamefont {T.}~\bibnamefont {Sj{\"o}strand}}, \bibinfo {author} {\bibfnamefont {S.}~\bibnamefont {Mrenna}}, \ and\ \bibinfo {author} {\bibfnamefont {P.~Z.}\ \bibnamefont {Skands}},\ }\href {\doibase 10.1088/1126-6708/2006/05/026} {\bibfield  {journal} {\bibinfo  {journal} {JHEP}\ }\textbf {\bibinfo {volume} {05}},\ \bibinfo {pages} {026} (\bibinfo {year} {2006})},\ \Eprint {http://arxiv.org/abs/hep-ph/0603175} {arXiv:hep-ph/0603175 [hep-ph]} \BibitemShut {NoStop}%
\bibitem [{\citenamefont {Sj{\"o}strand}\ \emph {et~al.}(2008)\citenamefont {Sj{\"o}strand}, \citenamefont {Mrenna},\ and\ \citenamefont {Skands}}]{Sjostrand:2007gs}%
  \BibitemOpen
  \bibfield  {author} {\bibinfo {author} {\bibfnamefont {T.}~\bibnamefont {Sj{\"o}strand}}, \bibinfo {author} {\bibfnamefont {S.}~\bibnamefont {Mrenna}}, \ and\ \bibinfo {author} {\bibfnamefont {P.~Z.}\ \bibnamefont {Skands}},\ }\href {\doibase 10.1016/j.cpc.2008.01.036} {\bibfield  {journal} {\bibinfo  {journal} {Comput. Phys. Commun.}\ }\textbf {\bibinfo {volume} {178}},\ \bibinfo {pages} {852} (\bibinfo {year} {2008})},\ \Eprint {http://arxiv.org/abs/0710.3820} {arXiv:0710.3820 [hep-ph]} \BibitemShut {NoStop}%
\bibitem [{\citenamefont {de~Favereau}\ \emph {et~al.}(2014)\citenamefont {de~Favereau}, \citenamefont {Delaere}, \citenamefont {Demin}, \citenamefont {Giammanco}, \citenamefont {Lemaitre}, \citenamefont {Mertens},\ and\ \citenamefont {Selvaggi}}]{deFavereau:2013fsa}%
  \BibitemOpen
  \bibfield  {author} {\bibinfo {author} {\bibfnamefont {J.}~\bibnamefont {de~Favereau}}, \bibinfo {author} {\bibfnamefont {C.}~\bibnamefont {Delaere}}, \bibinfo {author} {\bibfnamefont {P.}~\bibnamefont {Demin}}, \bibinfo {author} {\bibfnamefont {A.}~\bibnamefont {Giammanco}}, \bibinfo {author} {\bibfnamefont {V.}~\bibnamefont {Lemaitre}}, \bibinfo {author} {\bibfnamefont {A.}~\bibnamefont {Mertens}}, \ and\ \bibinfo {author} {\bibfnamefont {M.}~\bibnamefont {Selvaggi}} (\bibinfo {collaboration} {DELPHES 3}),\ }\href {\doibase 10.1007/JHEP02(2014)057} {\bibfield  {journal} {\bibinfo  {journal} {JHEP}\ }\textbf {\bibinfo {volume} {02}},\ \bibinfo {pages} {057} (\bibinfo {year} {2014})},\ \Eprint {http://arxiv.org/abs/1307.6346} {arXiv:1307.6346 [hep-ex]} \BibitemShut {NoStop}%
\bibitem [{\citenamefont {Cacciari}\ \emph {et~al.}(2012)\citenamefont {Cacciari}, \citenamefont {Salam},\ and\ \citenamefont {Soyez}}]{Cacciari:2011ma}%
  \BibitemOpen
  \bibfield  {author} {\bibinfo {author} {\bibfnamefont {M.}~\bibnamefont {Cacciari}}, \bibinfo {author} {\bibfnamefont {G.~P.}\ \bibnamefont {Salam}}, \ and\ \bibinfo {author} {\bibfnamefont {G.}~\bibnamefont {Soyez}},\ }\href {\doibase 10.1140/epjc/s10052-012-1896-2} {\bibfield  {journal} {\bibinfo  {journal} {Eur. Phys. J.}\ }\textbf {\bibinfo {volume} {C72}},\ \bibinfo {pages} {1896} (\bibinfo {year} {2012})},\ \Eprint {http://arxiv.org/abs/1111.6097} {arXiv:1111.6097 [hep-ph]} \BibitemShut {NoStop}%
\bibitem [{\citenamefont {Cacciari}\ and\ \citenamefont {Salam}(2006)}]{Cacciari:2005hq}%
  \BibitemOpen
  \bibfield  {author} {\bibinfo {author} {\bibfnamefont {M.}~\bibnamefont {Cacciari}}\ and\ \bibinfo {author} {\bibfnamefont {G.~P.}\ \bibnamefont {Salam}},\ }\href {\doibase 10.1016/j.physletb.2006.08.037} {\bibfield  {journal} {\bibinfo  {journal} {Phys. Lett.}\ }\textbf {\bibinfo {volume} {B641}},\ \bibinfo {pages} {57} (\bibinfo {year} {2006})},\ \Eprint {http://arxiv.org/abs/hep-ph/0512210} {arXiv:hep-ph/0512210 [hep-ph]} \BibitemShut {NoStop}%
\bibitem [{\citenamefont {Shih}(2021)}]{extraLHCOdataset}%
  \BibitemOpen
  \bibfield  {author} {\bibinfo {author} {\bibfnamefont {D.}~\bibnamefont {Shih}},\ }\href@noop {} {\enquote {\bibinfo {title} {Additional qcd background events for lhco2020 r\&d (signal region only)},}\ }\bibinfo {howpublished} {\url{https://zenodo.org/record/5759086}} (\bibinfo {year} {2021})\BibitemShut {NoStop}%
\bibitem [{\citenamefont {Thaler}\ and\ \citenamefont {Van~Tilburg}(2011)}]{Thaler:2010tr}%
  \BibitemOpen
  \bibfield  {author} {\bibinfo {author} {\bibfnamefont {J.}~\bibnamefont {Thaler}}\ and\ \bibinfo {author} {\bibfnamefont {K.}~\bibnamefont {Van~Tilburg}},\ }\href {\doibase 10.1007/JHEP03(2011)015} {\bibfield  {journal} {\bibinfo  {journal} {JHEP}\ }\textbf {\bibinfo {volume} {03}},\ \bibinfo {pages} {015} (\bibinfo {year} {2011})},\ \Eprint {http://arxiv.org/abs/1011.2268} {arXiv:1011.2268 [hep-ph]} \BibitemShut {NoStop}%
\bibitem [{\citenamefont {Thaler}\ and\ \citenamefont {Van~Tilburg}(2012)}]{Thaler:2011gf}%
  \BibitemOpen
  \bibfield  {author} {\bibinfo {author} {\bibfnamefont {J.}~\bibnamefont {Thaler}}\ and\ \bibinfo {author} {\bibfnamefont {K.}~\bibnamefont {Van~Tilburg}},\ }\href {\doibase 10.1007/JHEP02(2012)093} {\bibfield  {journal} {\bibinfo  {journal} {JHEP}\ }\textbf {\bibinfo {volume} {02}},\ \bibinfo {pages} {093} (\bibinfo {year} {2012})},\ \Eprint {http://arxiv.org/abs/1108.2701} {arXiv:1108.2701 [hep-ph]} \BibitemShut {NoStop}%
\bibitem [{\citenamefont {Komiske}\ \emph {et~al.}(2019)\citenamefont {Komiske}, \citenamefont {Metodiev},\ and\ \citenamefont {Thaler}}]{Komiske:2018cqr}%
  \BibitemOpen
  \bibfield  {author} {\bibinfo {author} {\bibfnamefont {P.~T.}\ \bibnamefont {Komiske}}, \bibinfo {author} {\bibfnamefont {E.~M.}\ \bibnamefont {Metodiev}}, \ and\ \bibinfo {author} {\bibfnamefont {J.}~\bibnamefont {Thaler}},\ }\href {\doibase 10.1007/JHEP01(2019)121} {\bibfield  {journal} {\bibinfo  {journal} {JHEP}\ }\textbf {\bibinfo {volume} {01}},\ \bibinfo {pages} {121} (\bibinfo {year} {2019})},\ \Eprint {http://arxiv.org/abs/1810.05165} {arXiv:1810.05165 [hep-ph]} \BibitemShut {NoStop}%
\bibitem [{\citenamefont {Aad}\ \emph {et~al.}(2020)\citenamefont {Aad} \emph {et~al.}}]{ATLAS:2020iwa}%
  \BibitemOpen
  \bibfield  {author} {\bibinfo {author} {\bibfnamefont {G.}~\bibnamefont {Aad}} \emph {et~al.} (\bibinfo {collaboration} {ATLAS}),\ }\href {\doibase 10.1103/PhysRevLett.125.131801} {\bibfield  {journal} {\bibinfo  {journal} {Phys. Rev. Lett.}\ }\textbf {\bibinfo {volume} {125}},\ \bibinfo {pages} {131801} (\bibinfo {year} {2020})},\ \Eprint {http://arxiv.org/abs/2005.02983} {arXiv:2005.02983 [hep-ex]} \BibitemShut {NoStop}%
\bibitem [{\citenamefont {Freytsis}\ \emph {et~al.}(2023)\citenamefont {Freytsis}, \citenamefont {Perelstein},\ and\ \citenamefont {San}}]{otherBDT}%
  \BibitemOpen
  \bibfield  {author} {\bibinfo {author} {\bibfnamefont {M.}~\bibnamefont {Freytsis}}, \bibinfo {author} {\bibfnamefont {M.}~\bibnamefont {Perelstein}}, \ and\ \bibinfo {author} {\bibfnamefont {Y.~C.}\ \bibnamefont {San}},\ }\href@noop {} {\  (\bibinfo {year} {2023})},\ \Eprint {http://arxiv.org/abs/2310.xxxxx} {arXiv:2310.xxxxx [hep-ph]} \BibitemShut {NoStop}%
\bibitem [{\citenamefont {Brun}\ and\ \citenamefont {Rademakers}(1997)}]{Brun:1997}%
  \BibitemOpen
  \bibfield  {author} {\bibinfo {author} {\bibfnamefont {R.}~\bibnamefont {Brun}}\ and\ \bibinfo {author} {\bibfnamefont {F.}~\bibnamefont {Rademakers}},\ }\href@noop {} {\bibfield  {journal} {\bibinfo  {journal} {Nucl. Inst. \& Meth. in Phys. Res. A}\ }\textbf {\bibinfo {volume} {389}},\ \bibinfo {pages} {81} (\bibinfo {year} {1997})},\ \bibinfo {note} {proceedings AIHENP'96 Workshop, Lausanne, Sep. 1996}\BibitemShut {NoStop}%
\bibitem [{\citenamefont {Akiba}\ \emph {et~al.}()\citenamefont {Akiba}, \citenamefont {Sano}, \citenamefont {Yanase}, \citenamefont {Ohta},\ and\ \citenamefont {Koyama}}]{Akiba:2019}%
  \BibitemOpen
  \bibfield  {author} {\bibinfo {author} {\bibfnamefont {T.}~\bibnamefont {Akiba}}, \bibinfo {author} {\bibfnamefont {S.}~\bibnamefont {Sano}}, \bibinfo {author} {\bibfnamefont {T.}~\bibnamefont {Yanase}}, \bibinfo {author} {\bibfnamefont {T.}~\bibnamefont {Ohta}}, \ and\ \bibinfo {author} {\bibfnamefont {M.}~\bibnamefont {Koyama}},\ }\href@noop {} {\enquote {\bibinfo {title} {Optuna: A next-generation hyperparameter optimization framework},}\ }\Eprint {http://arxiv.org/abs/arXiv:1907.10902} {arXiv:1907.10902} \BibitemShut {NoStop}%
\bibitem [{\citenamefont {Ng}(2004)}]{Ng:10.1145/1015330.1015435}%
  \BibitemOpen
  \bibfield  {author} {\bibinfo {author} {\bibfnamefont {A.~Y.}\ \bibnamefont {Ng}},\ }in\ \href {\doibase 10.1145/1015330.1015435} {\emph {\bibinfo {booktitle} {Proceedings of the Twenty-First International Conference on Machine Learning}}},\ \bibinfo {series and number} {ICML '04}\ (\bibinfo  {publisher} {Association for Computing Machinery},\ \bibinfo {address} {New York, NY, USA},\ \bibinfo {year} {2004})\ p.~\bibinfo {pages} {78}\BibitemShut {NoStop}%
\bibitem [{\citenamefont {Butter}\ \emph {et~al.}(2019)\citenamefont {Butter} \emph {et~al.}}]{Kasieczka:2019dbj}%
  \BibitemOpen
  \bibfield  {author} {\bibinfo {author} {\bibfnamefont {A.}~\bibnamefont {Butter}} \emph {et~al.},\ }\href {\doibase 10.21468/SciPostPhys.7.1.014} {\bibfield  {journal} {\bibinfo  {journal} {SciPost Phys.}\ }\textbf {\bibinfo {volume} {7}},\ \bibinfo {pages} {014} (\bibinfo {year} {2019})},\ \Eprint {http://arxiv.org/abs/1902.09914} {arXiv:1902.09914 [hep-ph]} \BibitemShut {NoStop}%
\end{thebibliography}%
\bibliographystyle{apsrev4-1}

\end{document}